\documentclass[fleqn,usenatbib]{mnras}

\newcommand{\tess}{\emph{TESS}}

\newcommand{\gaia}{\emph{Gaia}}
\newcommand{\ktwo}{\emph{K2}}

\newcommand{\kepler}{\emph{Kepler}}

\newcommand{\wise}{\emph{WISE}}

\newcommand{\teff}{\ensuremath{T_{\rm eff}}}

\newcommand{\primary}{HD~240779}
\newcommand{\secondary}{BD+10\,714B}

\usepackage{newtxtext,newtxmath}

\usepackage[T1]{fontenc}
\usepackage{ae,aecompl}

\usepackage{graphicx}	
\usepackage{amsmath}	
\usepackage{amssymb}	

\title[]{Planetesimals Around Stars with \tess\ (PAST):\\  I. Transient Dimming of a Binary Solar Analog at the End of the Planet Accretion Era}

\author[Gaidos et al.]{
Gaidos, E.,$^{1,2}$\thanks{E-mail: gaidos@hawaii.edu} T. Jacobs,$^{3}$ D. LaCourse,$^{4}$, A. Vanderburg,$^{5}$ S. Rappaport,$^{6}$ T. Berger,$^{7,2}$
\newauthor
L. Pearce,$^{5}$ A. W. Mann,$^{8}$ L. Weiss,$^{7,9,2}$ B. Fulton,$^{10}$ A. Behmard,$^{11}$ A. W. Howard,$^{10}$ 
\newauthor
M. Ansdell,$^{12,13}$, G. R. Ricker,$^{14}$ R. K. Vanderspek,$^{14}$ D. W. Latham,$^{15}$ S.Seager,$^{14,16,17}$
\newauthor
J. N. Winn$^{18}$ and J. M. Jenkins$^{19}$\\
$^{1}$Department of Earth Sciences, University of Hawai'i at M\={a}noa, Honolulu, HI  96822, USA\\
$^{2}$Kavli Institute for Theoretical Physics, UC Santa Barbara, Santa Barbara, CA 93106\\
$^{3}$12812 SE 69th Place Bellevue, WA 98006, USA\\
$^{4}$7507 52nd Place NE Marysville, WA 98270, USA\\
$^{5}$Department of Physics, Kavli Institute for Astrophysics and Space Research, M.I.T., Cambridge, MA 02139, USA\\
$^{6}$Department of Astronomy, The University of Texas at Austin, 2515 Speedway, Stop C1400, Austin, TX 78712, USA\\
$^{7}$Institute for Astronomy, University of Hawaii at M\={a}noa, Honolulu, HI 96822, USA\\
$^{8}$Department of Physics and Astronomy, University of North Carolina at Chapel Hill, Chapel Hill, NC 27599-3255, USA\\
$^{9}$Parrent Fellow\\
$^{10}$Department of Astronomy, California Institute of Technology, 1200 East California Boulevard, Pasadena, CA 91125, USA\\
$^{11}$Division of Geological and Planetary Sciences, California Institute of Technology, Pasadena, CA 91125, USA\\
$^{12}$Center for Integrative Planetary Science, University of California at Berkeley, Berkeley, CA 94720, USA\\
$^{13}$Department of Astronomy, University of California at Berkeley, Berkeley, CA 94720, USA\\
$^{14}$Department of Physics and Kavli Institute for Astrophysics and Space Research, MIT, Cambridge, MA 02139, USA\\
$^{15}$Center for Astrophysics | Harvard \& Smithsonian, 60 Garden Street, Cambridge, MA 02138, USA\\
$^{16}$Department of Earth, Atmospheric, and Planetary Sciences, MIT, Cambridge, MA 02139, USA\\
$^{17}$Department of Aeronautical and Astronautical Engineering, MIT, Cambridge, MA 02139, USA\\
$^{18}$Department of Astrophysical Sciences, Princeton University, Princeton, NJ 08544, USA\\
$^{19}$NASA Ames Research Center, Moffett Field, CA 94035, USA\\
}

\date{Accepted XXX. Received YYY; in original form ZZZ}

\pubyear{2019}

\hypersetup{draft}
\begin{document}
\label{firstpage}
\pagerange{\pageref{firstpage}--\pageref{lastpage}}
\maketitle

\begin{abstract}
We report detection of quasi-periodic (1.5-day) dimming of \primary, the solar-mass primary in a 5" visual binary (also TIC\,284730577), by the \emph{Transiting Exoplanet Survey Satellite}.  This dimming, as has been shown for other ``dipper" stars, is likely due to occultation by circumstellar dust.  The barycentric space motion, lithium abundance, rotation, and chromospheric emission of the stars in this system point to an age of $\approx$125~Myr, and possible membership in the AB Doradus moving group.  As such it occupies an important but poorly explored intermediate regime of stars with transient dimming between young stellar objects in star forming regions and main sequence stars, and between UX Orionis-type Ae/Be stars and M-type ``dippers".  \primary, but not its companion \secondary, has \emph{WISE}-detected excess infrared emission at 12 and 22 $\mu$m indicative of circumstellar dust.  We propose that infrared emission is produced by collisions of planetesimals during clearing of a residual disk at the end of rocky planet formation, and that quasi-periodic dimming is produced by the rapid disintegration of a $\gtrsim100$~km planetesimal near the silicate evaporation radius.  Further studies of this and similar systems will illuminate a poorly understood final phase of rocky planet formation like that which produced the inner Solar System.  
\end{abstract}

\begin{keywords}
stars: binaries -- kinematics and dynamics -- circumstellar matter -- planetary systems -- planet-star interactions -- planets \& satellites: protoplanetary disks  
\end{keywords}



\section{Introduction}
\label{sec:intro}

Observations of $\lesssim 10$~Myr-old Young Stellar Objects (YSOs) are a window into the past of the Solar System, as well as the thousands of planetary systems around main sequence stars revealed by space-based transit surveys, especially \kepler, and ground-based radial velocity surveys.  YSOs, including the low-mass T Tauri stars, were first identified by their variability \citep{Joy1945}, which remains a defining characteristic.  This variability is a result of the rotation of a spotted photosphere, flaring, episodic accretion, but also transient obscuration of the star by circumstellar matter (dust).  Variable Herbig Ae/Be stars were classified according to the behavior of their light curves \citep{Herbst1994}; ``Type III" variables, represented by the archetype UX Orionis, exhibit deep photometric minima that last weeks to months and occur episodically on timescales of years.  Such variables are readily identified and studied from the ground; changes in polarization and apparent color during dimming events indicate that dust, perhaps associated with the disk, are responsible for the obscuration \citep{Natta1997}.  AA~Tau, a T Tauri-type YSO, hinted at more diverse photometric variability among a wider range of stellar masses \citep{Bouvier1999}.  AA~Tau exhibited periodic (8.3~day) dimming until suddenly fading in 2013, behavior explained as obscuration by dust in a non-axisymmetric or ``funnel" accretion structure orbiting interior to the inner edge of a disk that is truncated and warped by the star's misaligned magnetic field \citep{Bouvier2014}.  The dramatic, evolving variability of another T Tauri-type binary KH\,15D is thought to be due to eclipses of one star by the precessing accretion disk of the other \citep[e.g.,][]{Herbst2010}.   

Space-based photometry of star-forming regions by \emph{CoRoT} and \emph{Spitzer}, unimpeded by Earth's rotation and atmosphere, led to identification of additional types of behavior, most notably ``dipper" stars with episodic transient dimming lasting for a fraction of day \citep{Alencar2010,Morales2011,Cody2014}.  The termination of the \kepler\ prime mission due to a failure of a second reaction wheel and advent of the two-wheeled \ktwo\ mission was a boon to YSO photometry since the latter observed several nearby star-forming regions: Taurus (1-3 Myr); $\rho$ Ophiucus ($\sim$3~Myr); Upper Scorpius ($\sim10$~Myr).  Only a minority of young stars observed by \ktwo\ exhibit dipping, but this could be a completeness (duty cycle) effect since these stars have only been observed for 70-80 days. 

As in the case of UX Orionis stars and AA Tau, dips are thought to be produced by dust, but the mechanism(s) and connection to the circumstellar disk and planet formation remain poorly understood.  Besides AA Tau-like accretion flows \citep{Bodman2017}, instabilities such as the Rossby wave instability could produce vertical structures in the disk that periodically occult the central star \citep{Stauffer2015,Ansdell2016}.  Disk-based scenarios predict that the disks of dipper stars should be highly inclined to the line of sight.  The outer regions, at least, of disks can be resolved by the Atacama Large Millimeter Array (ALMA) but the vast majority of these are \emph{not} highly inclined \citep{Ansdell2016b}.  Appeals to a warped inner disk may be applicable in some cases \citep{Loomis2017,Mayama2018}, but the limited work inside 1~AU on the brightest handful of objects with infrared interferometers shows moderately inclined inner disks as well.
\citep{Vural2014,Davies2018,Davies2018b}.  Alternative explanations for dips include dust lofted in disk winds, for which there is indirect support in other systems \citep{Varga2017,Fernandes2018,Ellerbroek2014}, gravitationally bound clumps of planetesimals \citep{Ansdell2016}, or disintegrating comet-like planetesimals \citep{Kennedy2017}. 

Transient dimming is rare but nevertheless does occur among main sequence stars that lack substantial disks.   These instances are important clues for understanding the full potential range of different circumstellar reservoirs and sources of dust.   Low-amplitude ($\lesssim$1\%) quasi-periodic dipping with Keplerian regularity but varying amplitude and shape has been explained by dust clouds emanating from ``evaporating" ultra-short period planets which by themselves are too small to produce a detectable transit signal \citep{Rappaport2012,Sanchis-Ojeda2015}.  Low-amplitude  dips with shapes consistent with transiting ``exocomets" have been identified in the light curves of several main sequence stars, none of which have detectable infrared excess indicative of disks \citep{Rappaport2018,Ansdell2019,Zieba2019}; the transiting objects could be larger analogs to the Sun-grazing comets discovered by the SOHO satellite \citep{Battams2017,Jones2018}.  The anomalous star KIC~8462852 \citep{Boyajian2016} may be the extreme member of a population of such stars \citep{Wyatt2018}.  Dips have also been detected in the light curve of at least one star that has evolved well beyond the main sequence, i.e., the white dwarf WD\,1145+017 \citep{Vanderburg2015}.  The transiting objects causing the dips are thought to be one or more evaporating, tidally disrupted planetesimals, a scenario consistent with observations infrared excess from circumstellar dust and accretion of heavy elements into the white dwarf atmosphere.

The transitional era between the YSO and mature stellar phases, populated by $\sim$100 Myr-old stars near the zero-age main sequence, could also illuminate the dipper phenomenon.  An example is RZ Piscium, an isolated, post-T Tauri star that is pre-main sequence but older than most dipper stars \citep[30-50 Myr][]{Grinin2010,Punzi2018}, has both UX Ori-like dimming and a substantial infrared excess from circumstellar dust \citep{deWit2013}. Its identification as a UX Ori-like variable was serendipitous since it lies well above the Galactic plane, far from any known star-forming region or known cluster, and lacks the defining attributes of a YSO.  

The \gaia\ all-sky astrometric survey will revolutionize our knowledge of young stars in the Solar neighborhood \citep{Lindegren2018}, in particular by identifying new members in young (co-) moving groups (YMGs) that are more kinematically than spatially clustered \citep{Faherty2018,Tang2019}.  The \emph{Transiting Exoplanet Survey Satellite} (\tess) is obtaining precise photometry of stars over 85\% of the sky with a cadence of $\leq$30\,min for at least 27 days \citep{Ricker2014}, permitting identification of dipper stars in these dispersed groups and even refining ages of the groups by gyrochronology \citep{Curtis2019}.  This \gaia-\tess\ synergy can be the basis of a broad, sensitive search for dipper stars with ages of tens to hundreds of Myr.  

Here we report that a \tess\ light curve of \primary, the primary component of a visual ($\rho = 5$") binary, reveals it to be a dipper star (Fig. \ref{fig:tess_lc}), and that other ground- and space-based observations suggest that it is a member of the $\approx$125 Myr-old AB Doradius Moving group (ABDMG).  In Section \ref{sec:observations} we present the data from \tess\ and other observations; in Section \ref{sec:analysis} we validate the source of the dip signal, determine stellar and orbital parameters, analyze the infrared excess due to circumstellar dust, and assess the age and moving group membership of the binary.  In Section \ref{sec:discussion} we summarize our findings and discuss the origin of the material responsible for the dipping and infrared excess, and compare this intriguing system to its closest published counterparts.  

\section{Observations and Data Reduction}
\label{sec:observations}

\subsection{\tess}

\begin{figure*}
	\includegraphics[width=\textwidth]{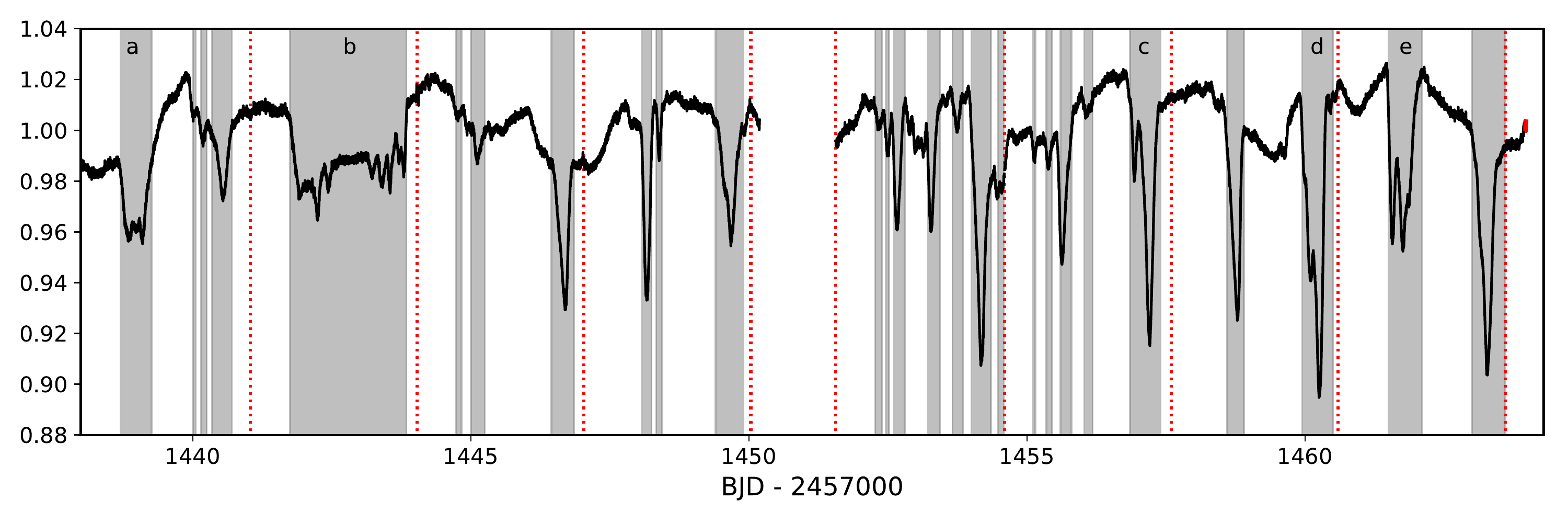}
    \caption{Normalized Sector 5 light curve from \tess\ 2-minute cadence data for an aperture centered on TIC~284730577/8.  Red points have quality flags set.  Vertical dashed red lines mark times where the spacecraft's hydrazine thrusters remove spin from the momentum wheels.  The interruption between 1450 and 1452 days is due to the spacecraft halting observations to transmit data at orbital perigee.  Grey regions are manually-selected dimming events or ``dips".}
    \label{fig:tess_lc}
\end{figure*}

\begin{figure}
	\includegraphics[width=\columnwidth]{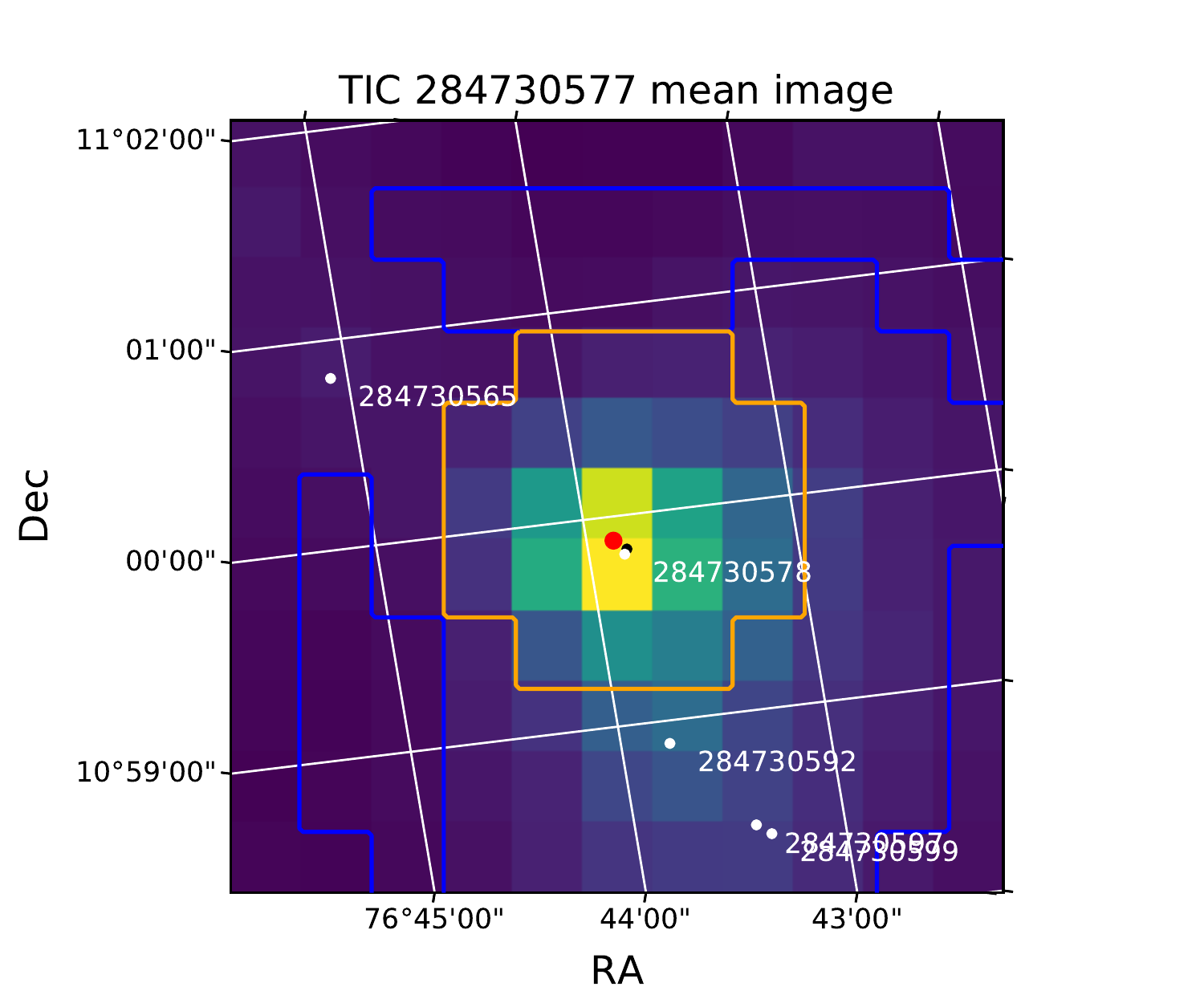}
    \caption{Mean \tess\ photometer image from Sector 5 observations of TIC 284730577 and 284730578.  The red point marks 284730577; other stars brighter than \tess\ magnitude $T = 15$ are marked by white points.  The black point is the flux centroid within the aperture.  Source and background apertures used to construct the light curve (Fig. \ref{fig:tess_lc}) are outlined by orange and blue lines, respectively.}
    \label{fig:tess_image}
\end{figure}

The binary consists of \primary\ and \secondary\ and was confirmed in \gaia\ Data Release 2 \citep{Lindegren2018} to be a physical pair at a distance of $95.2\pm0.4$\,pc (Table \ref{tab:params}).  The stars appear as sources 284730577 and 284730578 in the \tess\ Input Catalog (TIC) with \tess\ magnitudes $T=9.16$ and $T=9.98$ \citep{Stassun2018}.  The stars were observed by \tess\ during Sector 5 (orbits 17 and 18) from 15 November to 11 December 2018 in CCD detector 3 of Camera 1.  Cutouts (11 pixel $\times$11 pixel or 132"$\times$132") of the photometer images containing the two unresolved sources as well as source-free pixels for background subtraction were retained at 2-sec intervals and combined by onboard processes into a 2-min cadence sequence of images with a clipping algorithm to remove charged particle events prior to transmission to ground stations.  Figure \ref{fig:tess_image} is a mean image constructed from the entire sequence.  Besides the stars of interest, the only other stars of consequence (\tess\ magnitude $T < 15$, or at least 1\% as bright as the target stars) are plotted as white points in the figure.  Of particular interest is TIC~284730592 ($T \approx 12.0$), which may contribute a small amount of flux to the extracted light curve of TIC 284730578 and perhaps TIC 284730577, but cannot be the source of the dipping (see Section \ref{sec:source}).  A 18944-point light curve for each star was generated by the Science Processing Operations Center \citep[SPOC; ][]{Jenkins2016} using an optimized aperture centered on the position of each source, and subtracting a background signal from another aperture that avoids all sources in the cutout.  The light curves of the two stars are essentially identical since the aperture is more than an arc-minute across and includes both stars.  The light curve of the primary is shown in Fig. \ref{fig:tess_lc}.

\tess\ light curves are manually inspected by a team of citizen and academic scientists for notable phenomena, including dips.  The search is conducted using {\tt LcTools}\footnote{https://sites.google.com/a/lctools.net/lctools}, a free and publicly available software program that provides a set of applications for efficiently building and visually inspecting large numbers of light curves \citep{Kipping2015}. For more details on the {\tt LcTools} package and the visual survey methodology, see \citet{Rappaport2018}.  TIC~284730577/8 was identified as a light curve of interest on 3 March 2019.  The light curve and image data were retrieved from the Mikulski Archive for Space Telescopes (MAST).  

\subsection{Keck-1/HIRES}

We observed \primary\ and \secondary\ with the HIRES spectrometer \citep{Vogt1994}) at the W. M. Keck Observatory on UT 26 March 2019.  We used an exposure meter to stop the exposures after achieving a signal-to-noise ratio of 40 per pixel at the peak of the blaze function in the spectral order containing 550~nm light.  The spectral format and HIRES settings and reduction were identical to those used by the California Planet Search \citep[CPS;][]{Howard2010}. We used the C2 (14" $\times$ 0."86) entrance aperture of the decker mechanism and performed sky subtraction to reduce contamination from scattered light in the spectrograph.  The spectral coverage is from 3640 to 7990\AA. 

\section{Analysis}

\label{sec:analysis}

\subsection{Validation, Characterization and Source of Transient Dimming}
\label{sec:source}

\begin{figure}
	\includegraphics[width=\columnwidth]{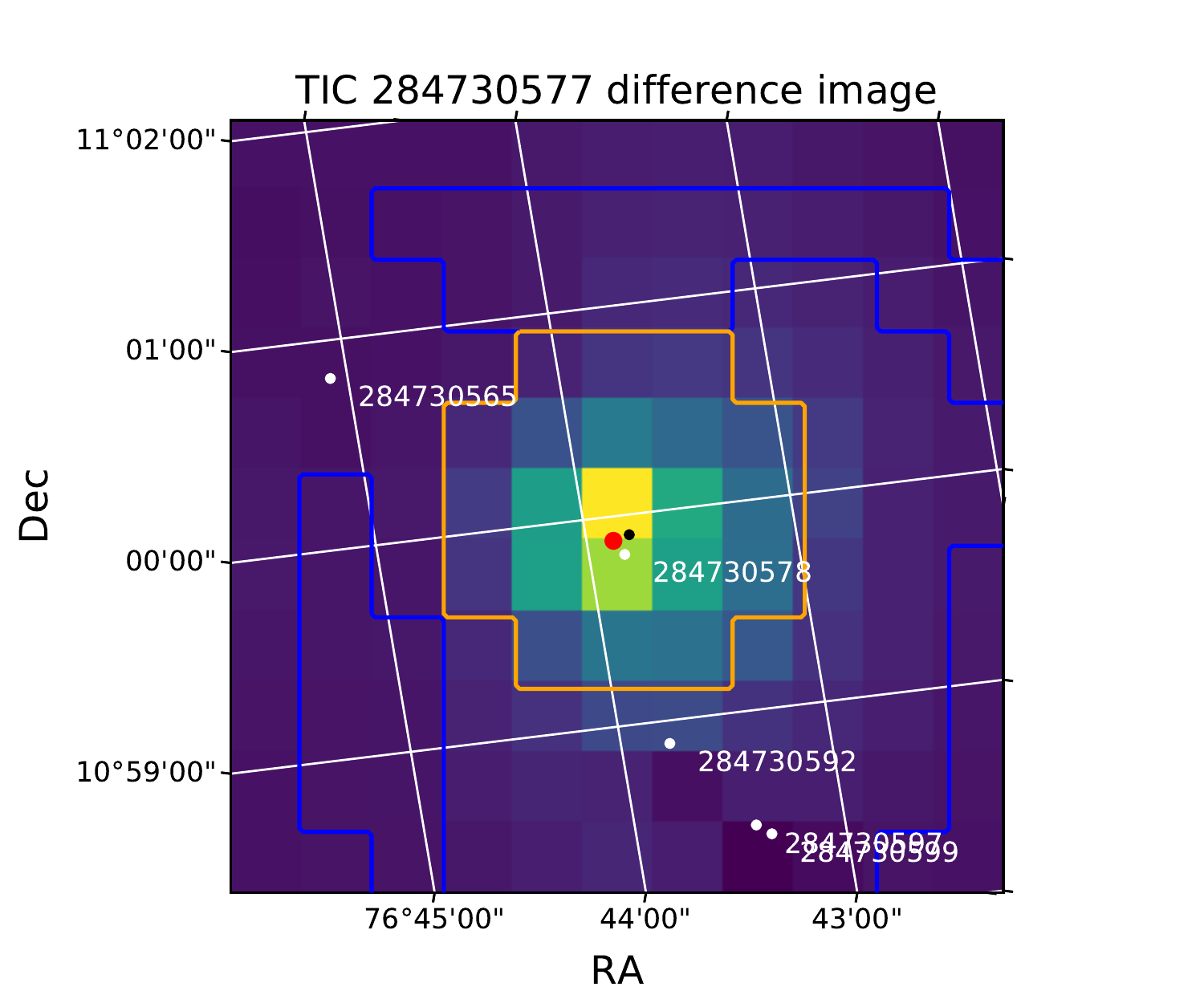}
    \caption{Difference image constructed by subtracting the mean of images during dips (demarcated by the grey regions in Fig. \ref{fig:tess_lc}) from the mean of image in an interval of equal duration before and after each of the dips.  The black dot marks the centroid location of the difference signal.  Its near coincidence with the mean source centroid (Fig. \ref{fig:tess_image}) demonstrates that the dipping signal is co-located with the binary star system.} 
    \label{fig:difference}
\end{figure}

Transient dimming events or ``dips" in the 2-min light curve of TIC 247306577/8 were identified by eye and highlighted in grey in Fig. \ref{fig:tess_lc}.  These do not correspond to times of the momentum dumps (vertical dotted red lines) indicating that they are unrelated.  \tess\ detector pixels subtend 21", meaning that confusion and mis-identification of a signal in an extracted light curve, particularly very faint signals typical for planetary transits, are possible.  In the case of the dips, the source cannot be fainter than $\approx$10\% (2.5 magnitudes) of the binary system.  The only nearby source that satisfies this requirement is TIC~2843730592 ($T \approx 12.0$), but this source is 1 arc-minute away and outside the aperture, so its flux contribution must be $\ll$10\%. A difference image constructed by subtracting the mean of images during dips from the mean of images in equal intervals before and after each dip (Fig. \ref{fig:difference}) shows that the ``missing" flux during dips is centered on the location of the system, demonstrating that this \emph{is} the source of the dimming signal.  The centroid location of the difference signal, marked by the black dot in Fig. \ref{fig:difference}, is only 0.2 pixels or 4" to the N of the mean source centroid (Fig. \ref{fig:tess_image}), a distance much smaller than the angular resolution of \tess\ and comparable to the binary separation.  Another potential source of false positives for dips is contamination of the background aperture by a variable or moving source, e.g., an asteroid.  We constructed a video using the cutouts which revealed no moving or transient sources.  Other distant sources can be the source of faint signals, e.g. by ``sticking" of charge from a bright variable star like an eclipsing binary \citep{Gaidos2017a} but these effects are unlikely to produce a 10\% signal in a bright star.

\begin{figure}
	\includegraphics[width=\columnwidth]{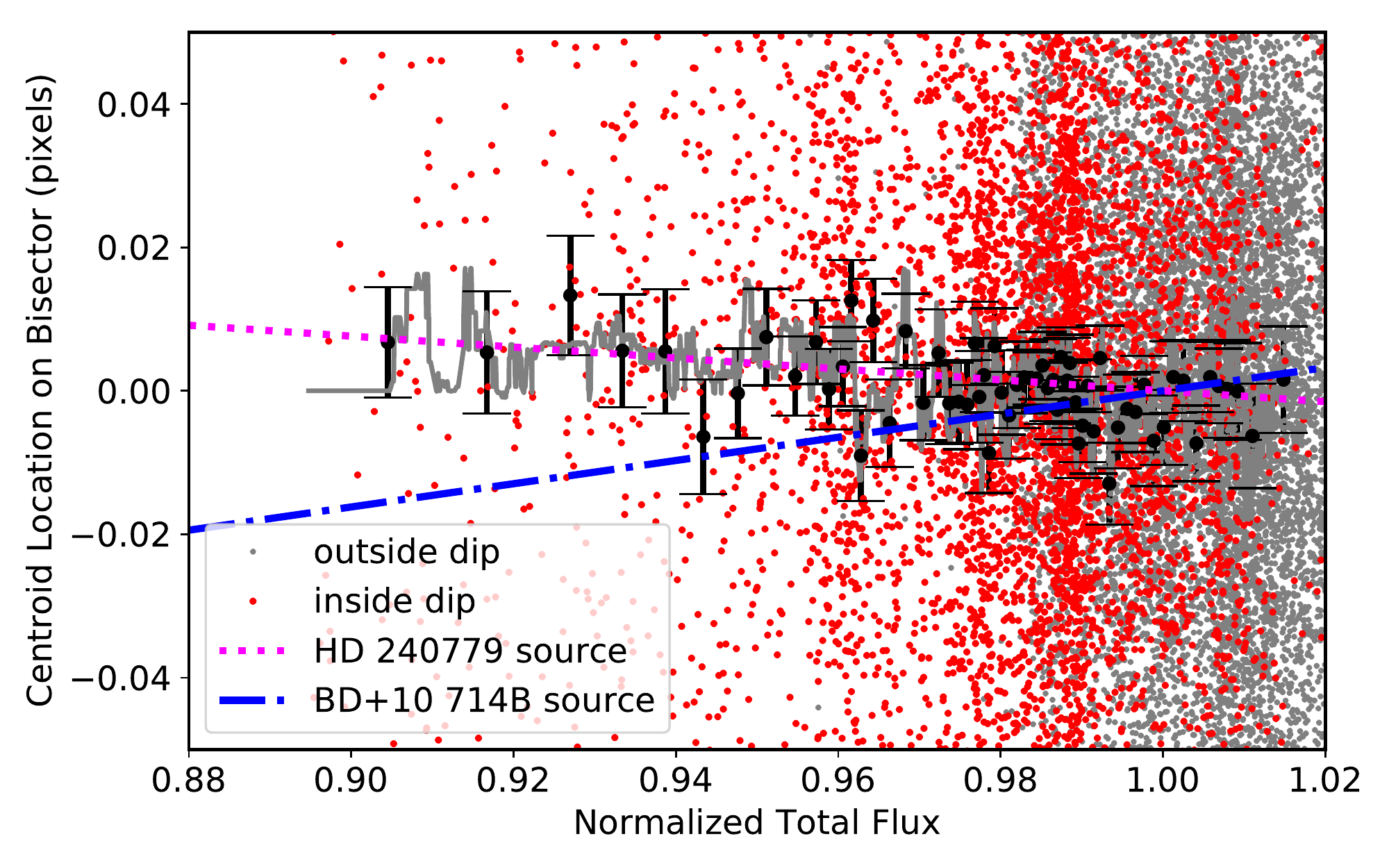}
    \caption{Relative position of the centroid of the unresolved \tess\ source containing flux from both \primary\ and \secondary, projected along the bisector between the two stars (positive towards \secondary) vs normalized flux.  The grey line is a running median filter with N=101, the black points are robust ($3\sigma$ clipping) means in bins of 101 points, with errors determined by 100 bootstrap with resampling.  The dotted magenta and dash-dot blue lines are the predicted trends based on a flux ratio in the \tess\ bandpass of 2.13.  Centroid motion is consistent with \primary\ as the source of the dips.}  
    \label{fig:centroid}
\end{figure}

We conclude that the binary system \primary+\secondary\ is responsible for the signal, but which star of the two (or both?) is involved? The components of this system are obviously not resolved by \tess.  However, the position of the unresolved source detected by \tess\ should lie at a flux-weighted position along a line between the two stars, and changes in the brightness of either star should induce a shift in the centroid $\gtrsim 0.01$~pixels which might be detected.  The magnitude and sign of the centroid shift depend on which star is dimming.  We projected the relative position of the moment-weighted centroid as supplied in the light curve file header ({\tt MOM\_CENT1} and {\tt MOM\_CENT2}) onto the bisector using the map projection coefficients supplied in the file.  The centroid drifts by several tenths of a pixel over the course of Sector 5 and this trend was removed with a $N=301$ running median filter.  Figure \ref{fig:centroid} shows the residual centroid shifts vs. the normalized flux.  Red points are those within the dips indicated by the grey zones in Fig. \ref{fig:tess_lc}; grey points are all others.  The grey line is running median of the points within dips and the black points are the robust means in 101-point bins.  A comparison with the expected motion (dotted magenta and dash-dotted blue lines) supports the primary star as the source of the dipping signal.  

\begin{figure}
	\includegraphics[width=\columnwidth]{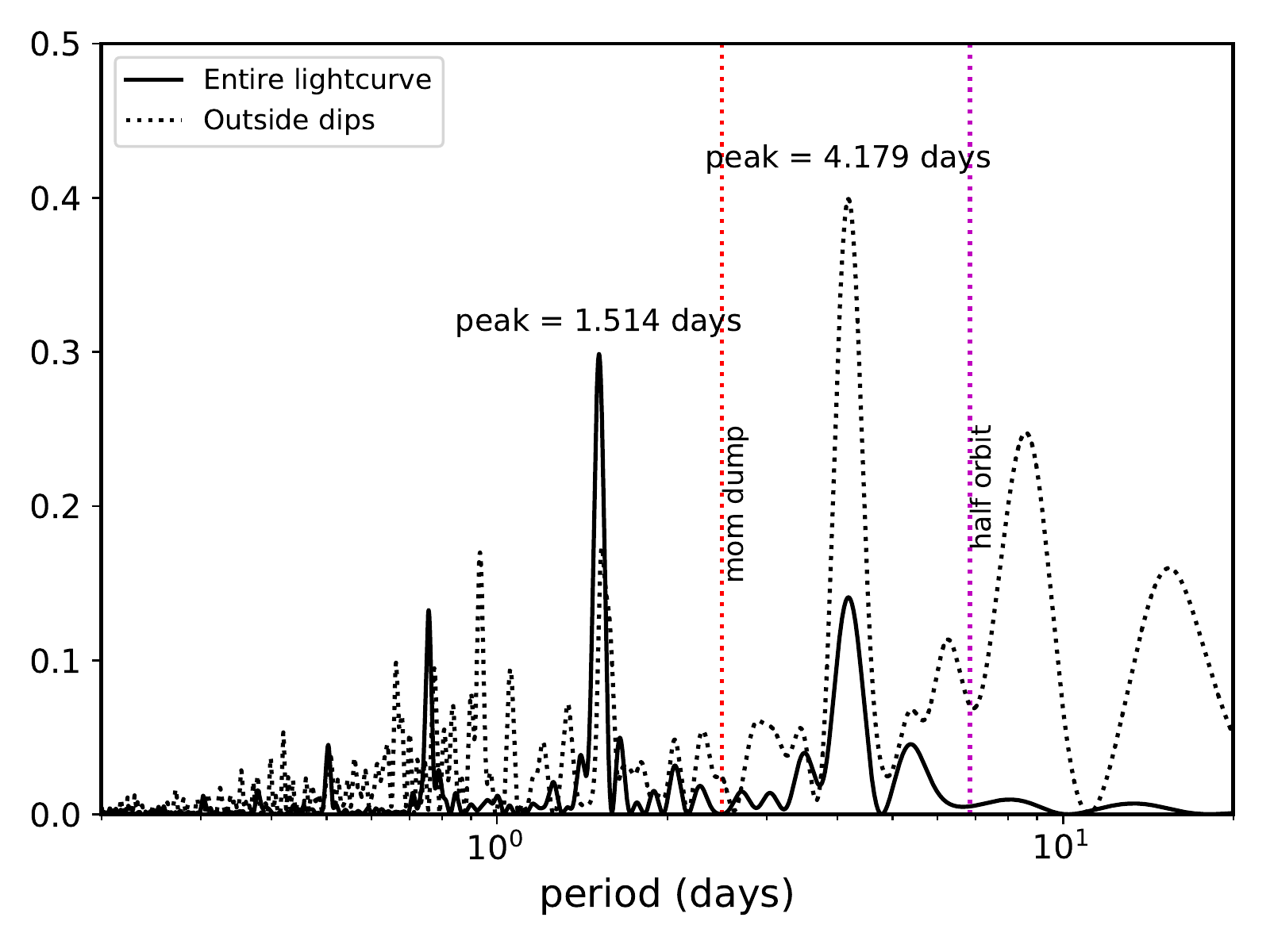}
    \caption{Lomb-Scargle periodogram of the \tess\ light curve, showing the peak at 1.51 days corresponding to dips.  The signal at $\approx$4.2 days could be due to rotational variability.  These signals are clearly distinct from any artifacts produced by removal of momentum from the reaction wheels and motion of the spacecraft during an orbit (vertical dotted lines).  The dotted curve is the periodogram constructed after removal of dipping intervals (grey bands in Fig. \ref{fig:tess_lc}). The signal at $\approx$0.8~days in the excised light curve could be a harmonic or alias of 1.51 days.}  
    \label{fig:periodogram}
\end{figure}

\begin{figure}
	\includegraphics[width=\columnwidth]{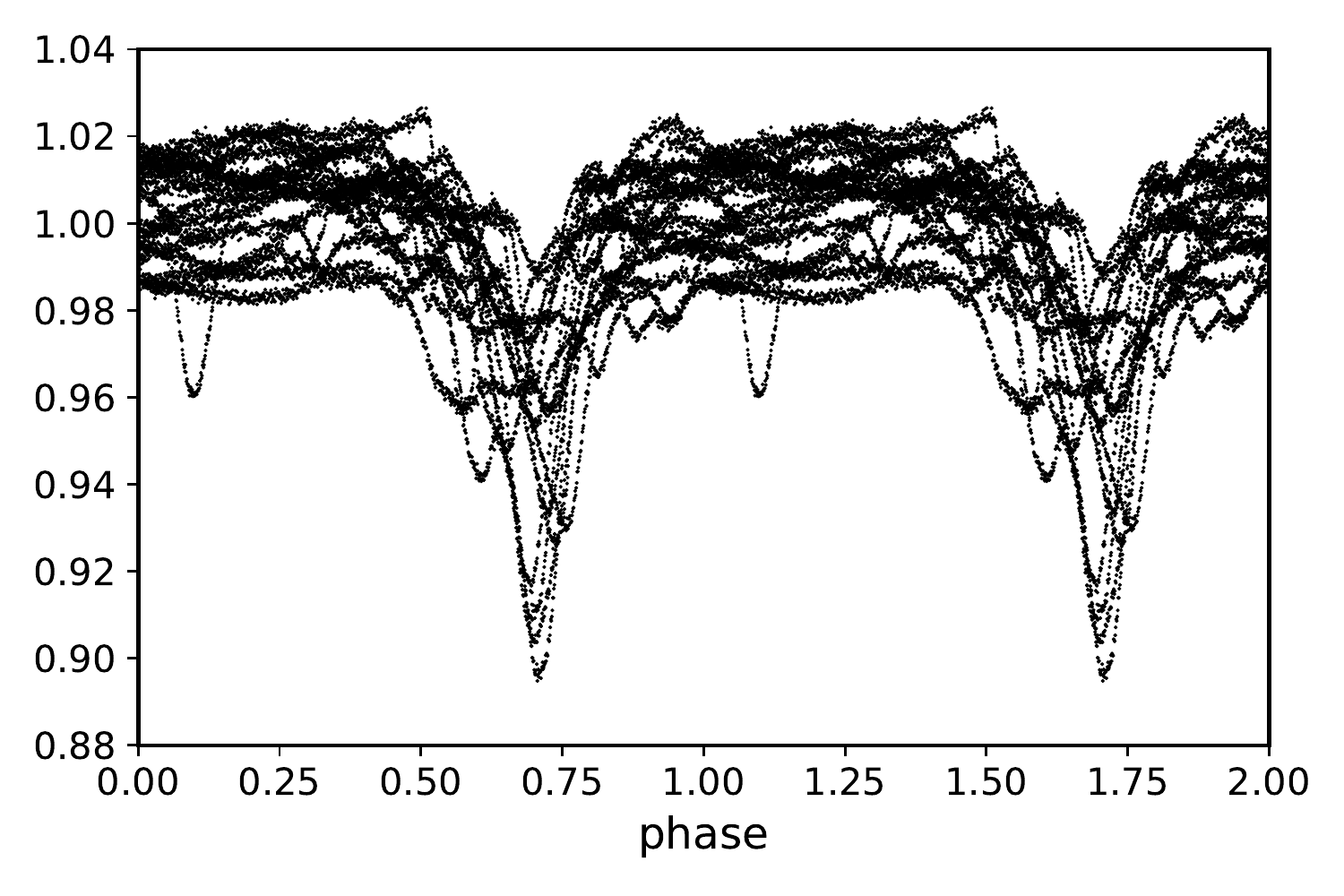}
    \caption{Normalized \tess\ light curve phased to 1.51 day period and repeated, showing the quasi-periodic nature of the signal.}  
    \label{fig:phase}
\end{figure}

Dimming events as deep as 10\% are superposed on irregular variability of about $\pm$2\%.  While most of these dips are much shorter than a day and have distinct minima, several (marked c, d, and e in Fig. \ref{fig:tess_lc}) appear to be double events, and two (marked a and b) are either single, long (1-2 day) events or composites of many unresolved events.    A Lomb-Scargle periodogram \citep{Scargle1982} of the entire light curve (solid line in Fig. \ref{fig:periodogram}), contains peaks at 1.51 and 4.2 days.  In a periodogram of the light curve after excision of the manually-identified dips, the 1.51 day-signal is suppressed and the 4.2 day signal is enhanced (dotted line in Fig. \ref{fig:periodogram}).  This indicates that the dipping is responsible for the 1.5-day signal.  Phased to this period (Fig. \ref{fig:phase}, the signal reveals its quasi-periodic nature, ruling out an eclipsing binary as the source.  We identify the 4.2-day with the rotational signal of the star (see Sec. \ref{sec:age}).

\begin{figure}
	\includegraphics[width=\columnwidth]{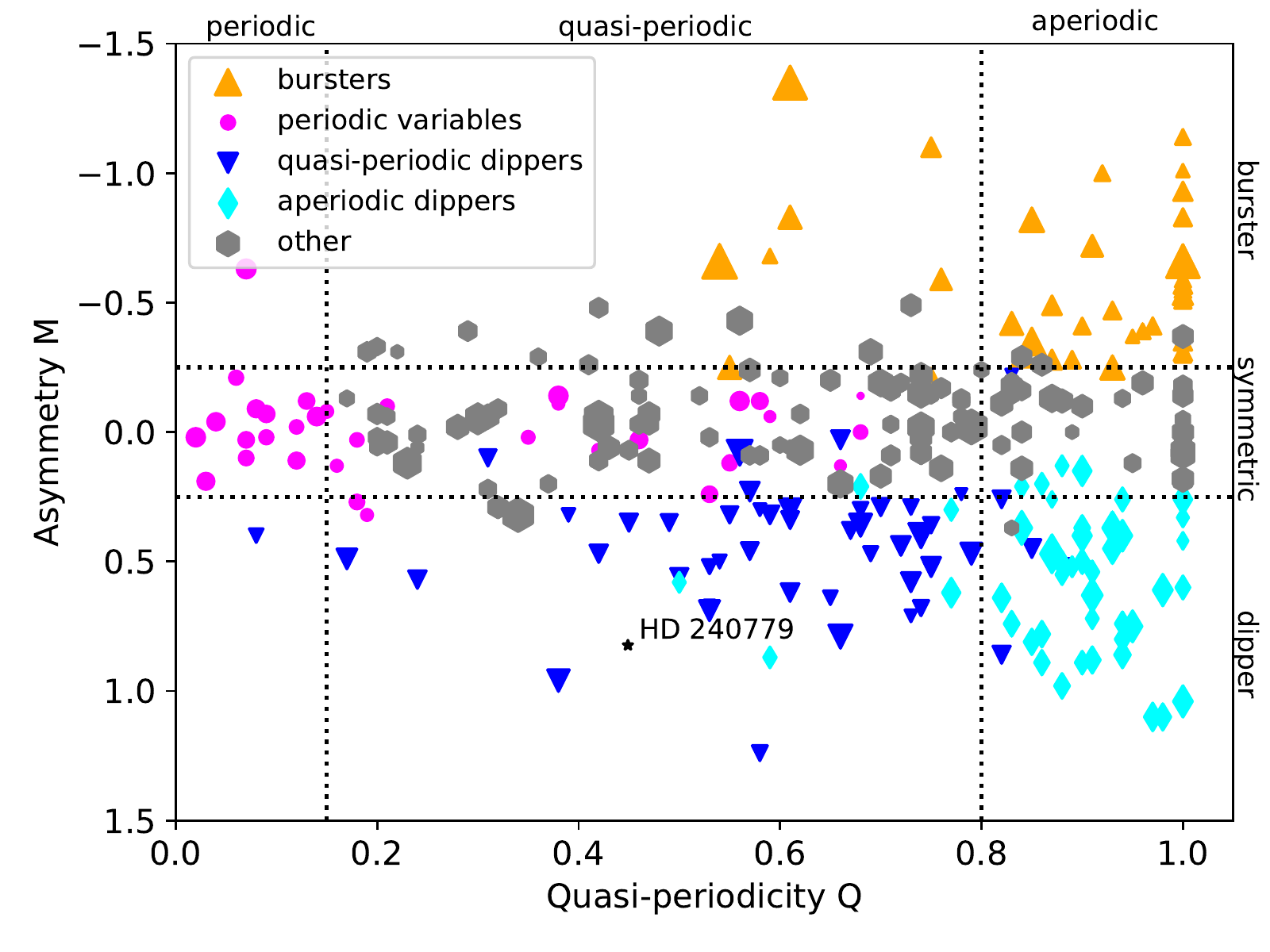}
    \caption{Asymmetry $Q$ and quasi-periodicity $M$ parameters of the \tess\ light curve of \primary\ (black star) compared to \ktwo\ light curves of variable stars in the Upper Scorpius and $\rho$ Ophiucus star-forming regions as determined by \citet{Cody2018}, using the definitions of $Q$ and $M$ defined in \citet{Cody2014}.  Marker sizes are colored according to the visual classification of \citet{Cody2018} and scaled by the square root of the variability amplitude.  The marker for \primary\ is also scaled to show its comparatively low variability.  The dotted lines mark the notional boundaries between the classes suggested by \citet{Cody2018}.}  
    \label{fig:cody}
\end{figure}

We quantified the overall morphology of the light curve to compare it with other dippers and variable stars using the asymmetry of the distribution of values, and the extent to which the lightcurve is quasi-periodic or aperiodic, that is to say the variability cannot be described by a single periodic signal.   To describe these properties, the asymmetry $M$ and quasi-periodicity $Q$ parameters defined by \citet{Cody2014} were calculated.  $M \in [-\infty,\infty]$ and a light curve with $M=0$ has a symmetric amplitude distribution, while one with $M=+1$ or $M=-1$ has many more negative- or positive- going points, respectively.  $Q \in [0,1]$; a light curve with $Q=0$ is perfectly periodic while having $Q=1$ is stochastic (no periodicity), and light curves with intermediate values of $Q$ are quasi-periodic, i.e. the periodic signal changes shape or amplitude, and/or there is additional variability at other timescales.  Dippers have $M>0$ while ``bursters" (stars with flaring and/or episodic accretion have $M<0$).  Rotational variables tend to have $M \approx 0$ and small $Q$.  The lightcurve of \primary\ + \secondary\ has $M=0.82$ and $Q=0.45$, placing it within the regime of quasi-periodic dippers in a $Q$-$M$ space (Fig. \ref{fig:cody}).  The overall variability amplitude (2\%) is modest compared to those of known dippers.  

\subsection{Stellar Parameters}
\label{sec:params}

HIRES spectra were extracted and processed using the procedures described in \citet{Howard2010}.  Spectra were matched to [synthetic/observational] library spectra using {\tt SpecMatch Synthetic and SpecMatch Empirical} \citep{Petigura2015,Petigura2017,Yee2017}.  The derived parameters are reported in Table \ref{tab:params}.   In addition, barycentric radial velocities (RVs) were derived and a search for additional sets of lines ruled out confused (spectroscopic binary) companions to a contrast ratio of 1\%.  Parameters were also estimated with the \teff\ and [Fe/H] from the HIRES spectrum and an absolute $K_s$-band magnitude using {\tt ISOCLASSIFY} \citep{Huber2017}.   These tools contain no special provisions for accurately determining ages or fitting young stars; SpecMatch Empirical uses a spectral library of middle-aged stars and there is no age prior for {\tt ISOCLASSIFY}. 

\begin{table}
\begin{center}
\caption{Stellar parameters} 
\label{tab:params}
\begin{tabular}{l | l | l  |  l}
& \multicolumn{1}{c}{{\tt SpecMatch}} & \multicolumn{1}{c}{{\tt SpecMatch}} & \multicolumn{1}{c}{{\tt Isoclassify}}\\
Parameter & \multicolumn{1}{c}{(Empirical)} & \multicolumn{1}{c}{(Synthetic} & \multicolumn{1}{c}{plus \gaia} \\
\hline
\multicolumn{4}{c}{\primary\ (TIC\,284730577)}\\
\hline
\teff\ [K] & 5791 (110) & 5793 (100) & 5780 (103)\\
$R_*$ [$R_{\odot}$] & 0.96 (0.10) & 0.97 (0.06) & 1.03 (0.01)\\
Fe/H & +0.04 (0.09) & +0.07 (0.06) & +0.03 (0.09) \\
log~g & -- & 4.58 (0.10) & 4.41 (0.03)\\
$M_*$ [$M_{\odot}$] & -- & 1.01 (0.04) & 1.00 (0.05) \\
log~age [yr] & -- & 9.36 (0.42) & 9.63 (0.26)\\  
RV [km s$^{-1}$] & +14.86 (0.10) & & \\
$v \sin i$ [km s$^{-1}$] & \multicolumn{3}{c}{11.0 (1.0)}\\
d [pc] & \multicolumn{3}{c}{94.7 (0.5)}\\
\hline
\multicolumn{4}{c}{\secondary\ (TIC\,284730578)}\\
\hline
\teff\ & 5099 (110) & 5193 (100) & 5070 (80)\\
$R_*$ [$R_{\odot}$] & 0.78 (0.10) & 0.83 (0.04) & 0.79 (0.01)\\
Fe/H & +0.04 (0.09) & +0.15 (0.06) & 0.03 (0.08)\\
log~g & -- & 4.66 (0.10) & 4.55 (0.03)\\
$M_*$ [$M_{\odot}$] & -- & 0.88 (0.03) & 0.83 (0.03)\\
log~age [yr] & -- & 9.62 (0.42) & 9.75 (0.3)\\
RV [km s$^{-1}$] & +16.04 (0.10) & &\\
$v \sin i$ [km s$^{-1}$] & \multicolumn{3}{c}{6.2 (1.0)}\\
d [pc] & \multicolumn{3}{c}{95.6 (0.6)}\\
\hline
\end{tabular}
\end{center}
\end{table}

\subsection{Orbit Fitting}
\label{sec:orbit}

\begin{figure}
	\includegraphics[width=\columnwidth]{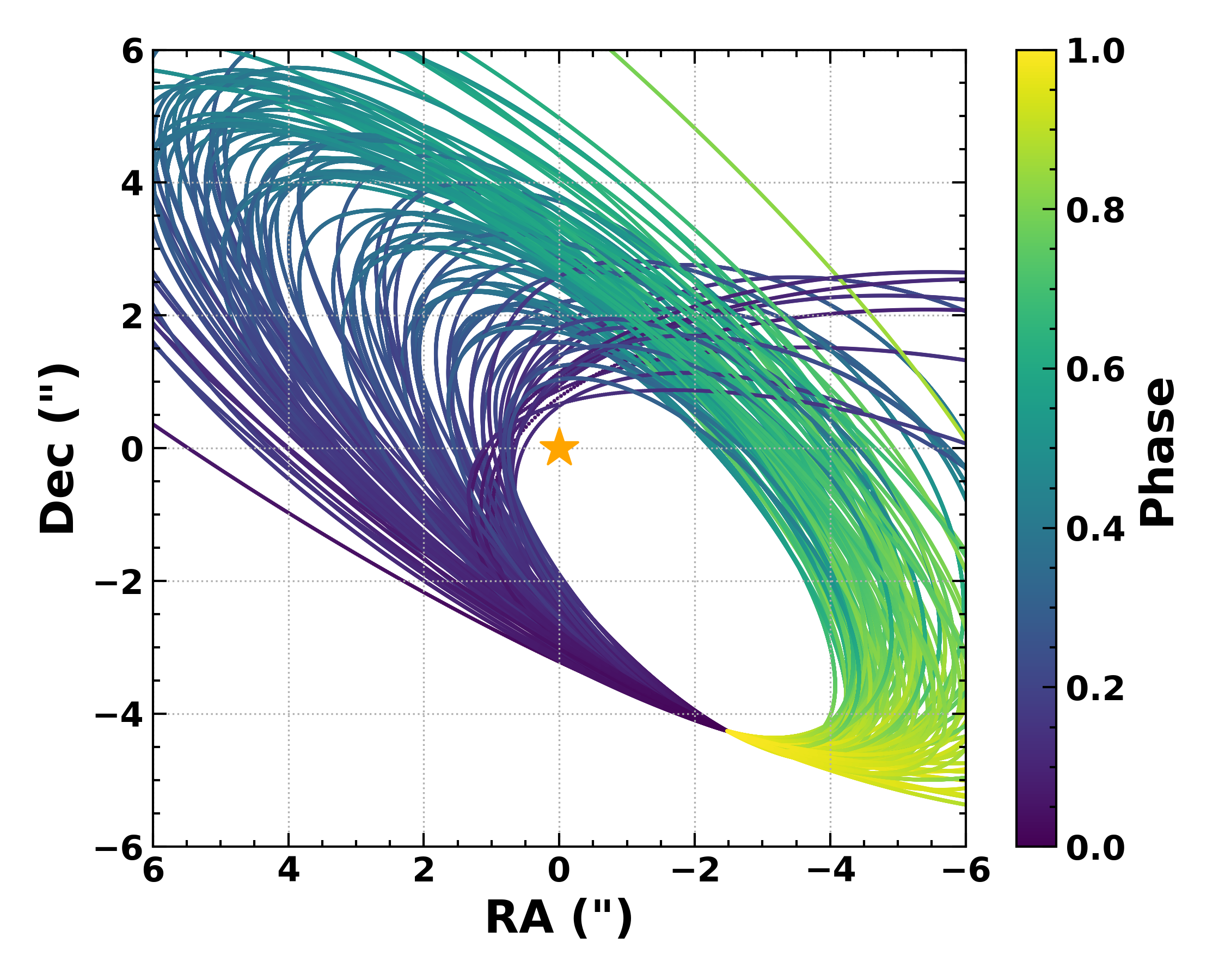}
    \caption{Subset of possible orbits of \secondary\ relative to \primary\ (marked as the star) constrained by \gaia\ astrometry, an RV measurement for each star, and spectroscopically-based mass estimates.}  
    \label{fig:orbits}
\end{figure}

\begin{figure}
	\includegraphics[width=\columnwidth]{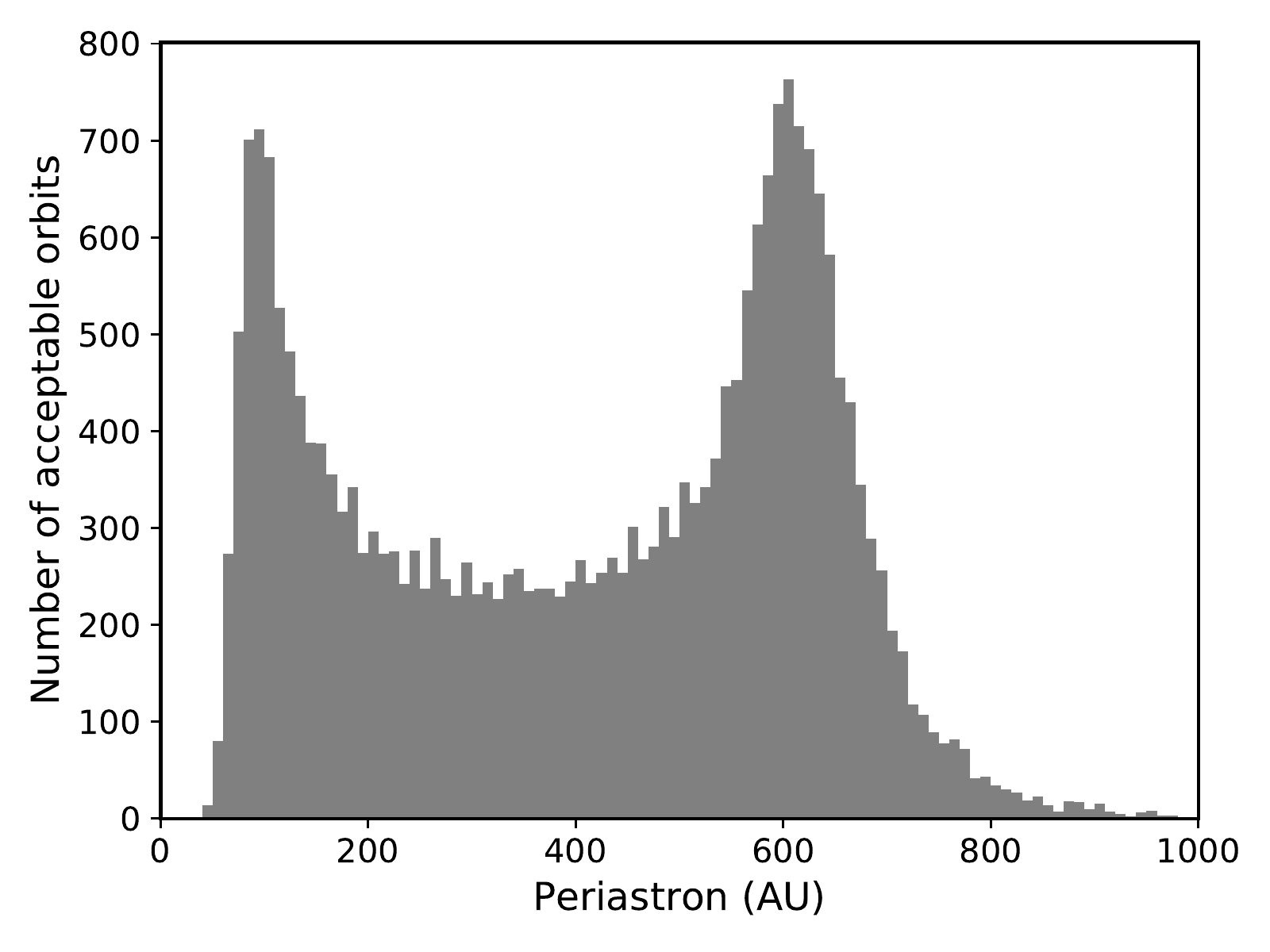}
    \caption{Distribution of possible periastra of 25535 orbits of \primary\ + \secondary\ that satisfy constraints imposed by \gaia\ and an RV measurement.}  
    \label{fig:periastra}
\end{figure}

\gaia\ parallaxes and proper motions of the two components indubitably show them to be at the same distance (within errors) and nearly the same sky-projected motion.  With a projected separation of $\approx 500$~AU and hence orbital period of many thousands of years, obtaining a true astrometric orbit is not possible.  However, a family of possible orbits can be identified and the range of possible orbital parameters constrained using current positions and motions.  Acceptable orbits  were identified using the {\tt LOFTI} implementation of the Orbits For the Impatient code ({\tt OFTI}) \citep{Blunt2017,Pearce2019}, using the constraints of the current \gaia\ separation and proper motions of each component, the RVs at a single epoch $\approx$3.5~yr after \gaia\ DR2 (2015.5) (Table \ref{tab:params}) and an inferred total system mass (the empirical values from Table \ref{tab:params}).  An initial semi-major axis ``guess" of 220~AU and a uniform prior for eccentricity were used.  Figure \ref{fig:orbits} plots 100 randomly-selected sky-projected orbits from the sample, colored by phase relative to the current epoch.  Figure \ref{fig:periastra} plots the periastron distribution of 25535 accepted orbits as an indicator of the maximum potential gravitational interaction between the stars.  The distribution is very broad, with peaks near 90~AU and 600~AU.  The possibility that the periastron is $<100$ AU has implications for the structure of the disk around \primary\ and the mechanism responsible for the transient dimming (Sec. \ref{sec:discussion}).          

\subsection{Infrared Excess}
\label{sec:infrared}

\begin{figure}
	\includegraphics[width=\columnwidth]{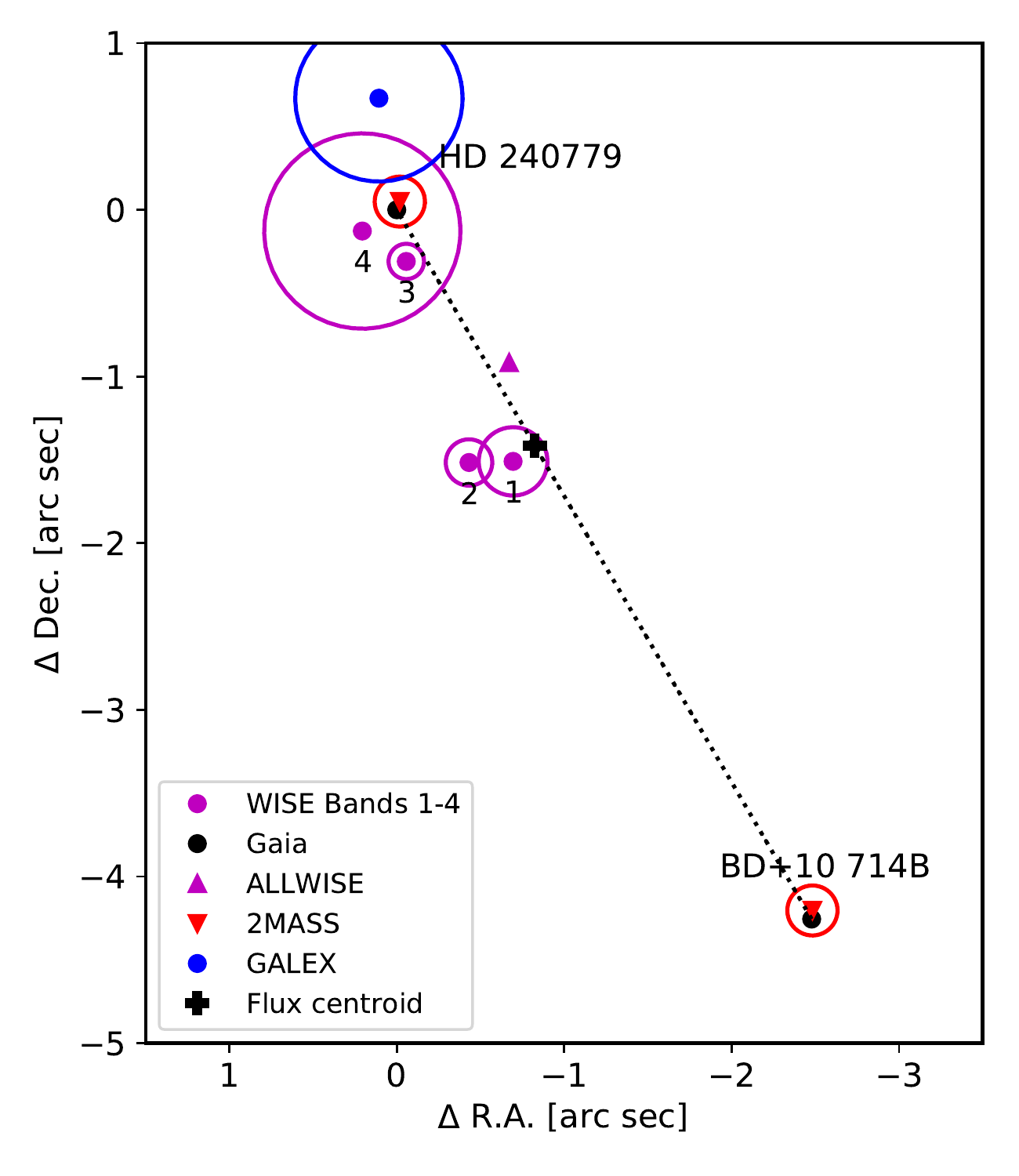}
    \caption{Epoch 2015.5 relative positions of \primary\ and \secondary\ as measured by \gaia\ along with infrared sources from 2MASS (red) and WISE (magenta).  Infrared source locations have been adjusted to 2015.5 using \gaia\ proper motions.  The circles are scaled to the errors in position.}  
    \label{fig:sources}
\end{figure}

The system is resolved by the 2MASS survey and the source positions (adjusted to epoch 2015.5 using \gaia\ DR2 proper motions) correspond within errors to the \gaia\ DR2 positions of the stars (Fig. \ref{fig:sources}).  Images in the individual \wise\ 1-4 passbands were downloaded from the NASA/IPAC Infrared Science Archive and the source associated with this system inspected.  In no bandpass was the binary resolved, although in the highest resolution image (W1, 6.1") the source is elongated in the NE-SW direction of the binary.  The centroid of each source were determined by fitting elliptical Gaussians as implemented by {\tt DAOSTARFINDER} in the {\tt PHOTUTILS} package in Python.  These are plotted in Fig. \ref{fig:sources} along with the ALLWISE source position.  While the ALLWISE \citep{Cutri2013} and W1 and W2 positions correspond to the expected flux centroid calculated from $K_s$ magnitudes, the W3 (12$\mu$m) and W4 (22$\mu$m) sources fall on the position of the \primary, indicating that it is the source of the detected infrared excess. 

\begin{figure}
	\includegraphics[width=\columnwidth]{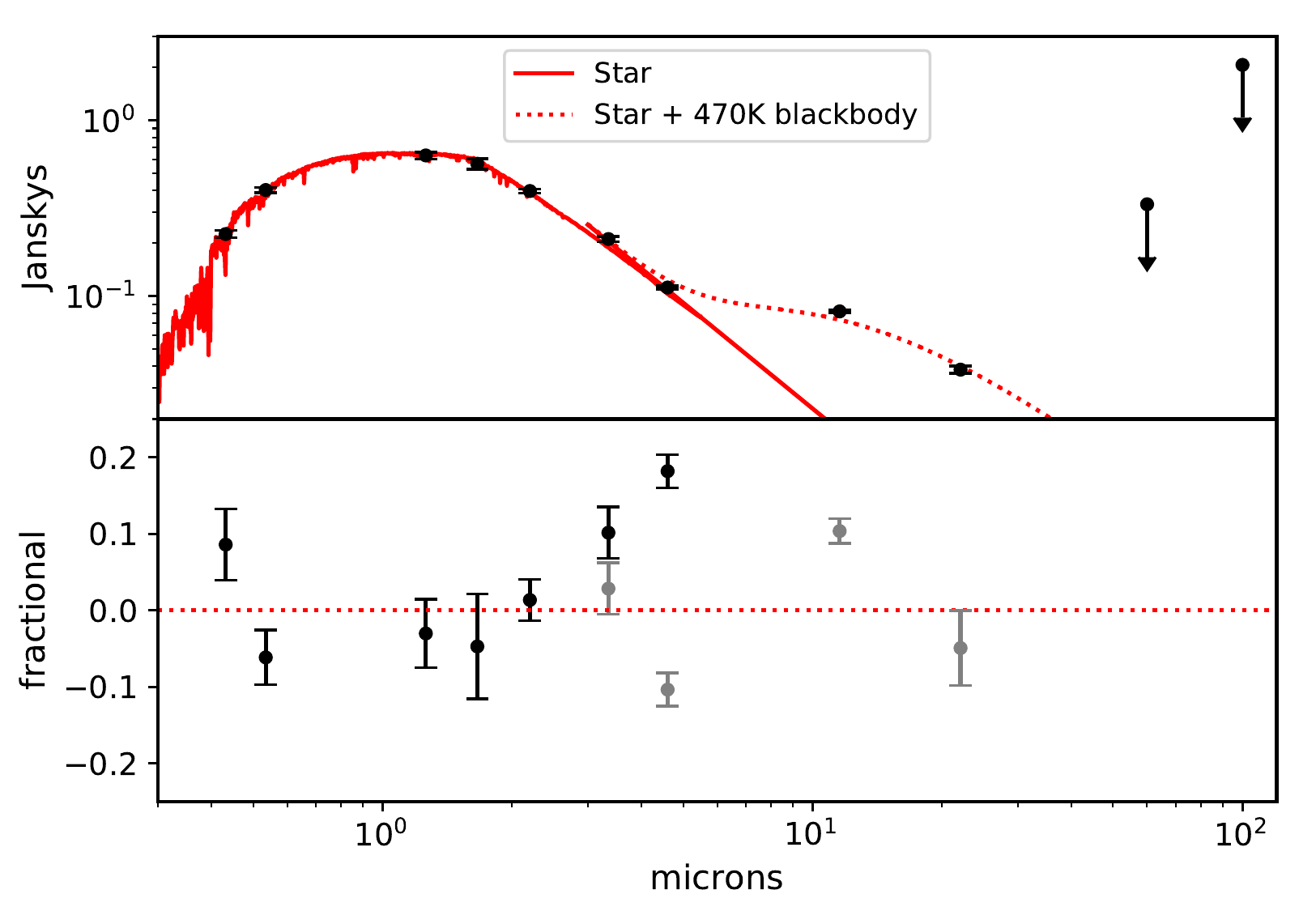}
    \caption{Top: Spectral energy distribution of \primary\ using \emph{Tycho}, 2MASS, and \emph{WISE} photometry.  The solid red line is a {\tt PHOENIX} model atmosphere with parameters (\teff=5800K, $\log g = 4.5$, [Fe/H] = 0, [$\alpha$/Fe] = 0).  The dashed red line includes a 470K blackbody, i.e. due to circumstellar dust.  Bottom: difference between the SED and stellar atmosphere model (black points) and star + blackbody (grey points).}
    \label{fig:sed}
\end{figure}

Figure \ref{fig:sed} plots flux density vs. effective wavelength for \primary\ based on resolved photometry from \emph{Tycho} ($B_T$ and $V_T$), 2MASS ($JHK_S$), and \emph{WISE} (channels W1-4 with effective wavelengths of 3.4, 4.6, 12, and 22 $\mu$m).  W1 and W2 flux densities were proportioned between \primary\ and \secondary\ according to their relative brightness in $K_s$-band.  The system is unresolved in AAVSO Photometric All Sky Survey (APASS) and was not observed by \emph{Spitzer}, \emph{Herschel}, or at (sub)mm wavelengths.  Effective wavelengths and zero points were drawn from \citet{Mann2015b}.  A smoothed model PHOENIX spectrum from \citep{Husser2013} with \teff=5800K, [Fe/H]=0, $\alpha$/Fe=0, $\log g =4.5$ with a best-fit normalization, plus a $\lambda^{-4}$ Rayleigh-Jeans extension, is shown.  The fit is markedly improved ($\chi^2 = 7.5$ $\nu = 3$ degrees of freedom) by the inclusion of reddening ($E_{B-V}=0.11$).    Extinction maps \citep[e.g.,][]{Green2018} are not properly calibrated within $\sim$100~pc, but negligible reddening is expected with the ``Local Bubble" ($d \ll 100$~pc).  However, we caution against interpreting this as evidence for circumstellar reddening, since the star is variable in the optical and the \emph{Tycho} and 2MASS observations were at different epochs.  

Excess emission is obvious in the \emph{WISE} W3 (12$\mu$m) and W4 (22$\mu$m) channels, but less evident in the W1 (3.4$\mu$m) and W2 (4.6$\mu$m) channels.  This is consistent with the W1 and W2 photocenters coinciding with the $K_s$-band centroid (presumably dominated by photosphere flux) between the two stars (Fig. \ref{fig:sources}).  The excess is approximately consistent with a 470K blackbody with a solid angle of $2.8 \times 10^{-17}$~Sr or about 170 times that of the stellar photosphere.  A blackbody model added to the Rayleigh-Jeans extension of the stellar model is plotted as a dotted red line in the top panel of Fig. \ref{fig:sed} and the difference between the observations and model plotted as the grey points in the bottom panel.  Unsurprisingly, the data are not well fit by a single-temperature blackbody, which may mean a range of dust temperatures or silicate emission near 10$\mu$m.  The inferred fractional luminosity, assuming a blackbody spectrum, is $L_{\rm dust}/L_{*} = 2 \times 10^{-3}$.      

\subsection{Young Moving Group Membership and Age}
\label{sec:age}

\begin{figure}
	\includegraphics[width=\columnwidth]{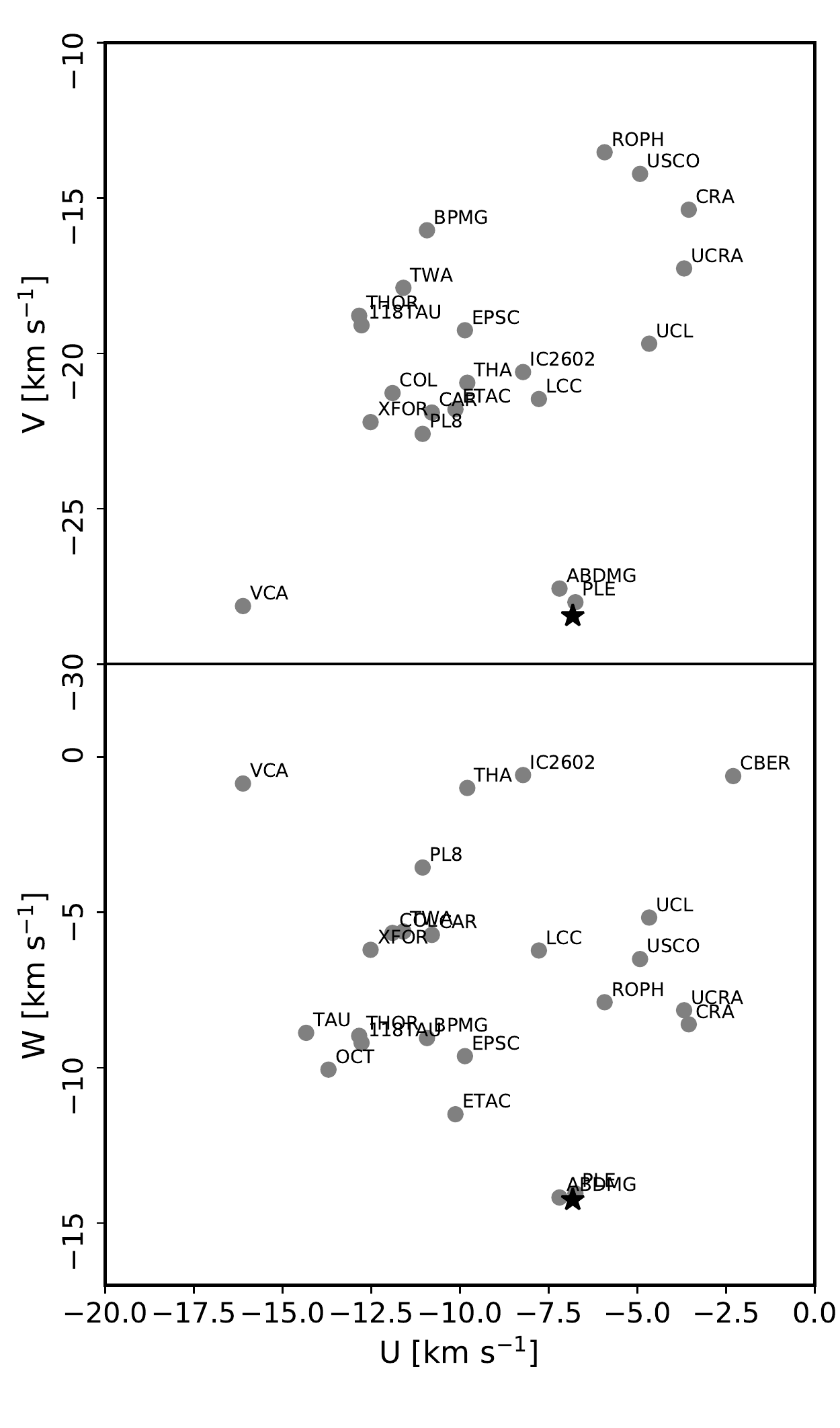}
    \caption{$UVW$ Galactic space motion of the \primary\ + \secondary\ system relative to the motions of clusters and moving groups of young stars in the {\tt Banyan $\Sigma$} database  \citep{Gagne2018}.}  
    \label{fig:uvw}
\end{figure}

Unlike nearly all previously identified dipper stars, the \primary\ + \secondary\ system is not a known or proposed member of a nearby star-forming region.  However, precise parallaxes and proper motions from \gaia\ plus the HIRES radial velocity were used to calculate Galactic $UVW$ space motion and show a kinematic link with the $\approx$125 Myr-old Pleiades cluster and AB Doradus moving group (ABDMG, Fig \ref{fig:uvw}.   Color-magnitude diagrams \citep{Luhman2005,Barenfeld2013} and a ``traceback" age of 125~Myr \citep{McCarthy2014} suggest a kinematic and chronological connection to the Pleiades \citep{Ortega2007}.

The Bayesian {\tt Banyan $\Sigma$} algorithm, which uses both $UVW$ and space coordinates $XYZ$ \citep{Gagne2018}, indicates a 44\% probability of membership in the ABDMG and negligible probability for every other cluster and moving group in the database, including the Pleiades; the remainder probability is assigned to the field.  This system lies 60~pc from the center of the Pleiades, well outside the 13~pc tidal radius of the cluster \citep{Adams2001}.  Likewise, this system is more distant than typical ABDMG members, although the dispersion of that group is larger and more poorly known.

\begin{figure}
	\includegraphics[width=\columnwidth]{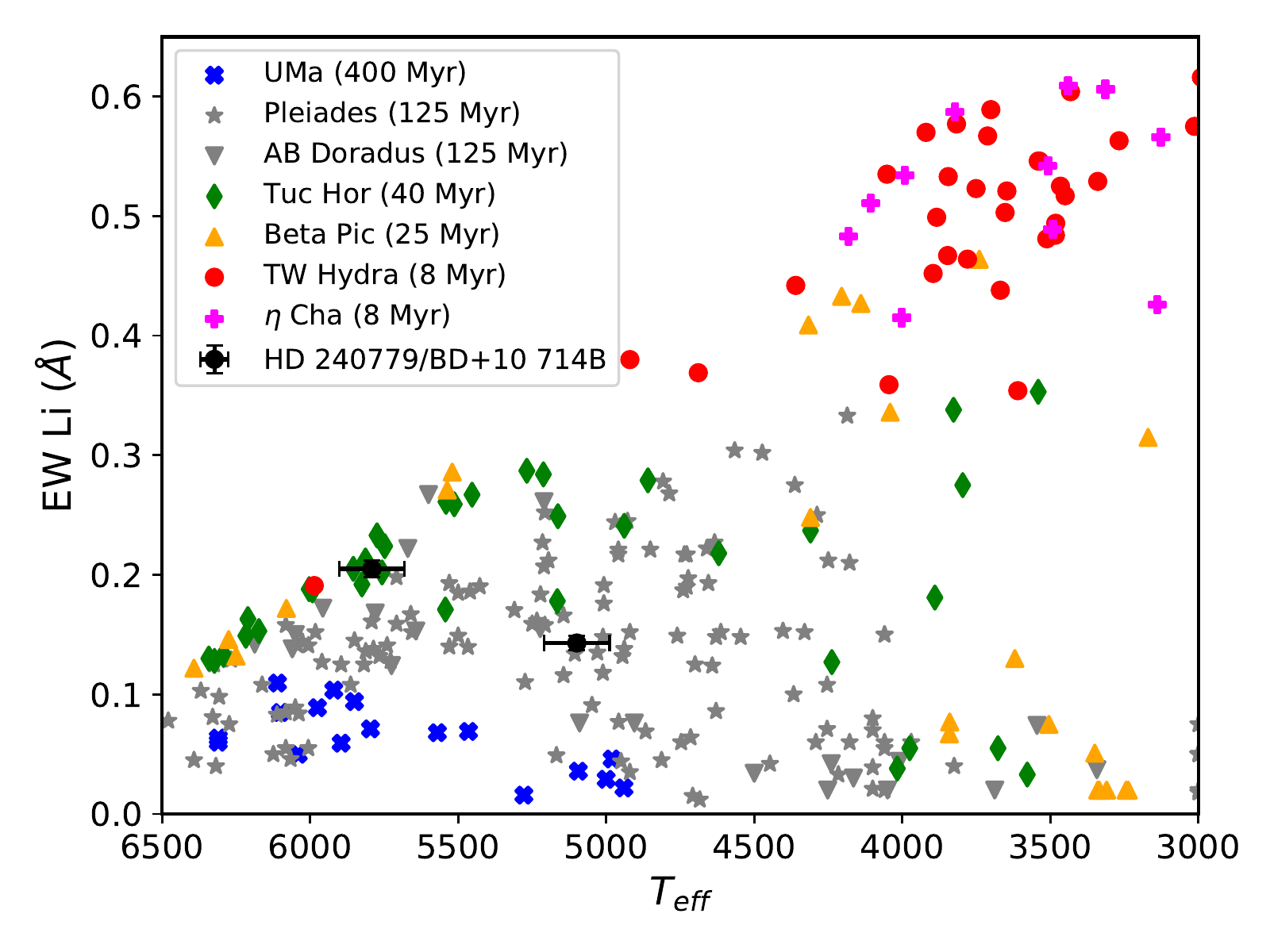}
    \caption{Li EW measured in spectra of \primary\ and \secondary, compared to those in nearby young moving groups, showing consistency with AB Dor and Ple age stars but not much younger or older groups.}  
    \label{fig:lithium}
\end{figure}

Lithium abundance is a widely-used chronometer for solar-type stars.  Because Li is destroyed as it is carried deep into the stellar interior by the surface convection zones of main sequence stars, the abundance of the element is observed to decrease with time, and more rapidly with cooler stars with deeper convection zones.  We measured the equivalent width (EW) of the 6708\AA\ doublet of Li~I following a procedure similar to that described in \citet{Berger2018b}. However, instead of using a Levenberg-Marquardt fit to measure the lithium equivalent width (EW), we utilized {\tt ROBOSPECT} \citep{Waters2013} which provided a better fit to the Li features while also ignoring the contribution of the blended Fe I line at 6707.441 \AA. From the spectrum of \primary, we measured an EW$_{\mathrm{Li}}$ = 143 $\pm$ 6 m\AA\ and 205 $\pm$ 7 m\AA\ for \secondary.  Figure \ref{fig:lithium} compares these values with those of members of nearby young moving groups compiled from the literature.  Measurements for the Pleiades are from measurements and compilation by \citet{Bouvier2017}, for Ursa Majoris from \citet{Ammler-von-Eiff2009}, and all the remaining are from \citet{Mentuch2008}.  The abundance of Li in the \primary\ + \secondary\ system falls well within the range spanned by the Pleiades and ABDMG, but not younger or older groups (Fig. \ref{fig:lithium}).

Another widely used chronometer for solar-type stars is rotation.  If a star is sufficiently spotted, rotation of the photosphere produces a sinusoidal-like signal in a light curve which can be identified by a periodogram or autocorrelation function.  

\begin{figure}
	\includegraphics[width=\columnwidth]{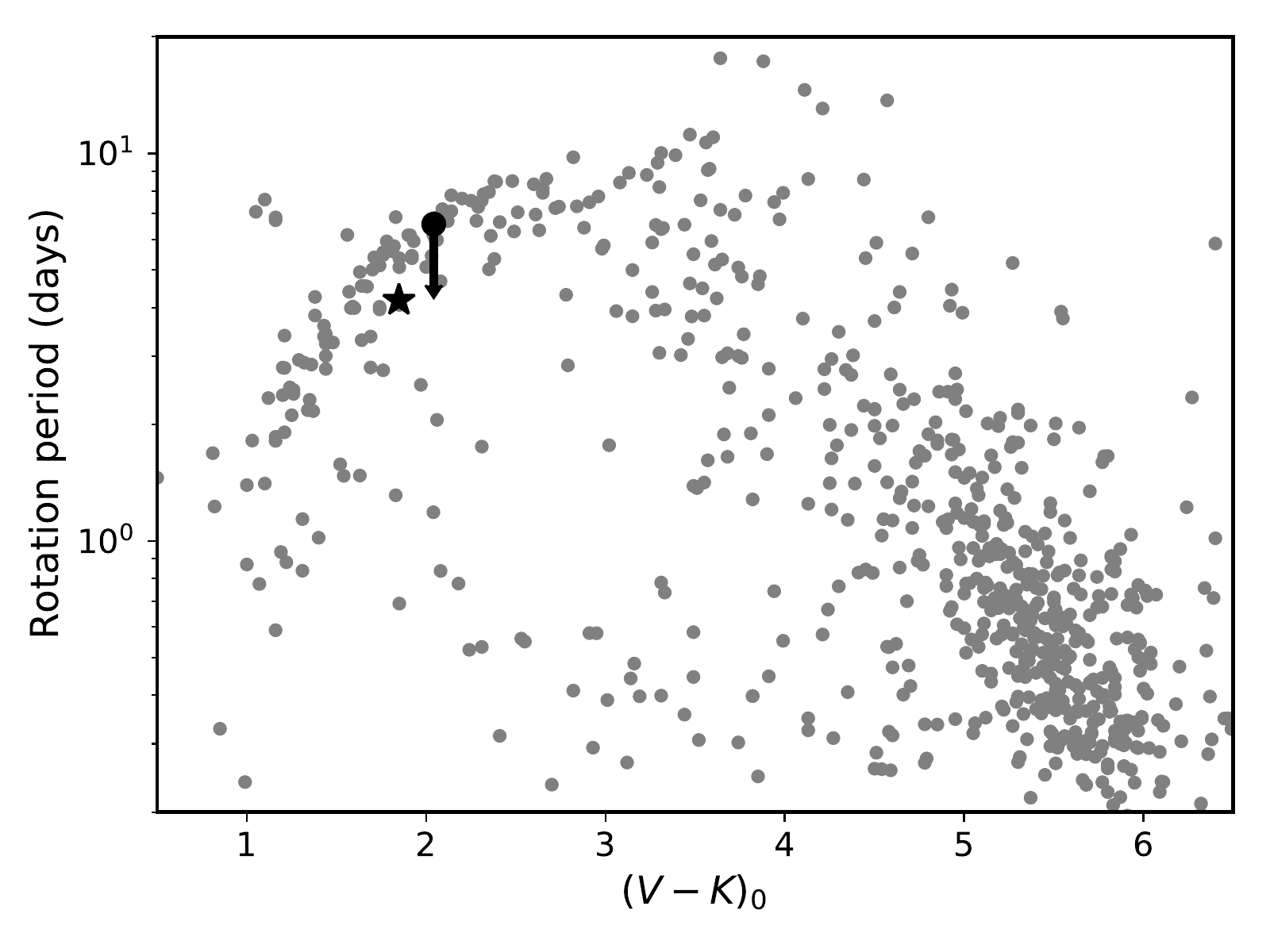}
    \caption{Possible rotation periods for \primary\ (black star) and \secondary\ (upper limit) compared to the distribution in 125 Myr Pleiades stars as measured from \ktwo\ data by \citet{Rebull2016}.}  
    \label{fig:rotation}
\end{figure}

 Although an autocorrelation analysis (not shown) does not confirm the 4.2-day signal, it is consistent with the 4.4-day upper limit imposed by the $v \sin i$ and $R_*$ of \primary\ (Tab. \ref{tab:params}), which is assumed to dominate the signal.  Those of \secondary\ suggest a rotation period no longer than 6.6 days.  Both these periods are consistent with a Pleiades-like age (Fig. \ref{fig:rotation}).

\begin{figure}
	\includegraphics[width=\columnwidth]{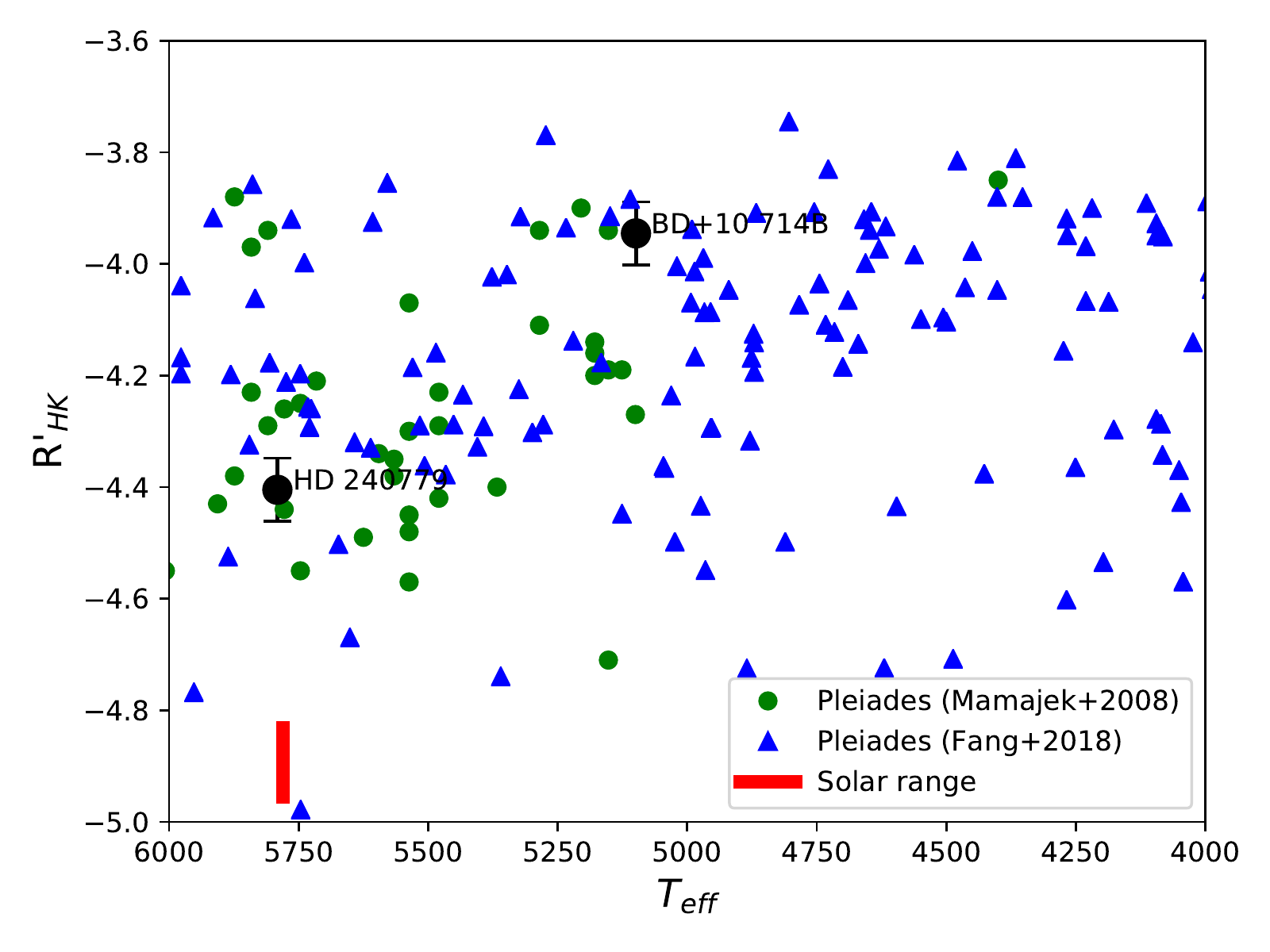}
    \caption{Chromospheric emission index $R'_{HK}$ of compared to that of Pleiades stars from \citet{Mamajek2008} and \citet{Fang2018}.  The red bar is the range spanned by the quiet and active Sun.}  
    \label{fig:cahk}
\end{figure}

The magnetic activity produced by a rotation-driven dynamo will heat the upper atmosphere of a star, producing emission in the H and K lines of singly-ionized calcium, as well as many lines in the far and near ultraviolet wavelengths.  The Ca II HK emission index $R'_{HK}$ was calculated using the conversion constants for the Mount Wilson $S$ index formula obtained in \citet{Wright2004}, and the photosphere contribution (which must be subtracted) using the formula with $B-V$ color from \citet{Noyes1984}.  Values for $\log R'_{HK}$ of -4.40 and -3.95 were determined for \primary\ and \secondary, respectively.  In Fig. \ref{fig:cahk} these are compared to values from Pleiades stars from \citet{Mamajek2008}, with a conversion from $B-V$ color to \teff\ from \citet{Pecaut2013}, and from \citet{Fang2018}.   The dipper system is far more magnetically active than the Sun (red bar in Fig. \ref{fig:cahk}) and comparable in activity to Pleiades members.  

\begin{figure}
	\includegraphics[width=\columnwidth]{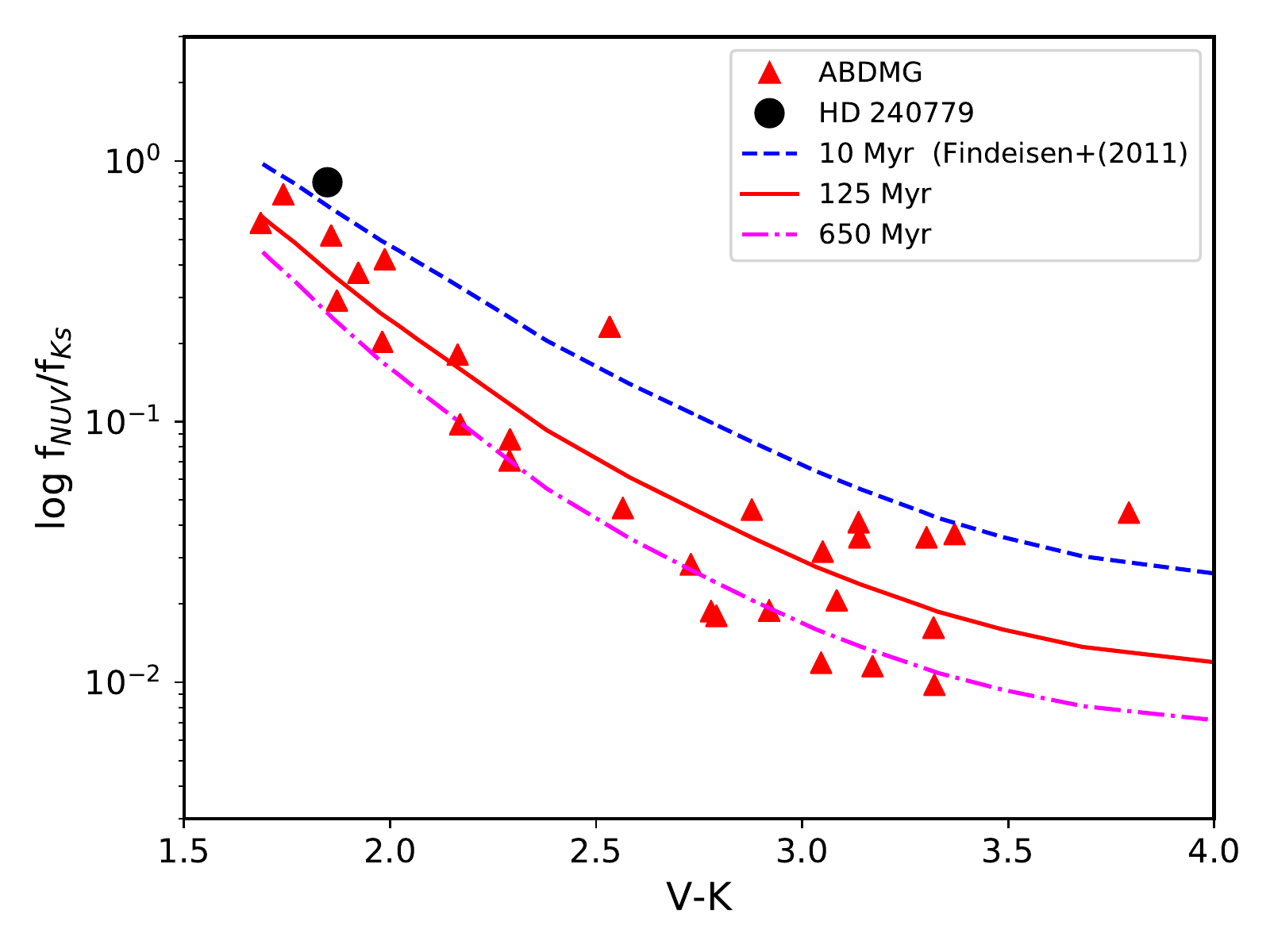}
    \caption{NUV to $K_s$-band flux density of \primary\ vs $V-K_s$ color compared to empirical relations for from  \citet{Findeisen2016} for three stellar ages (curves) and to confirmed AB Doradus Moving Group members (red triangles, J. Gagn\'{e}, personal communication).}
    \label{fig:galex}
\end{figure}

The system was detected in both the near ultraviolet (NUV; 1770-2730\AA) and far ultraviolet (FUV; 1350-1780\AA) channels of the \emph{Galex} space telescope \citep{Bianchi2017}.  The NUV source coincides with the location of \primary\ (Fig. \ref{fig:sources} and the FUV-NUV source offset is 0."65, not significant compared to the positional uncertainty (0."47).  The ratio of the flux density in the NUV over that in the $K_s$ band is similar to that of known ABDMG members (Fig. \ref{fig:galex}) and the empirical relations of \citet{Findeisen2016} for young stars.

\section{Summary and Discussion}
\label{sec:discussion}

Like the photometric survey satellites that preceded it, \tess\ is expanding our knowledge of the scope of stellar variability, including that produced from occultation by circumstellar dust.  \tess\ 2-min cadence observations of the solar-type primary \primary\ of a double star yielded a 27-day light curve containing multiple dimming events, each lasting a fraction of a day with attenuation up to 10\%, and often spaced by 1.5 days.  The asymmetry and quasi-periodicity of the light curve, as quantified by the $M$ and $Q$ parameters of \citet{Cody2014}, distinguish it from those of rotational variables and classify it as a quasi-periodic dipper, although some anomalous young, very fast-rotating M dwarfs also have light curves that are highly periodic but asymmetric in time and amplitude for reasons that are not understood \citep{Stauffer2018,Zhan2019}.  

The binary system's barycentric space motion matches closely with that of the 125~Myr Pleiades and the AB Doradus Young Moving Group, but its position clearly rules out membership in the Pleiades.  Measurements of Li abundances, Ca II HK line emission, and NUV emission are all consistent with a 125 Myr age and thus this system could belong to a spatially extended ABDMG, or a larger, kinematically connected population that is coeval with the Pleiades and ABDMG.  The metallicity of \primary\ and \secondary\ found by analysis of a high-resolution spectrum (Table \ref{tab:params}) are slightly but not significantly ($0.1 \pm 0.1$) above that of the AB Doradus moving group \citep{McCarthy2014}.  But if this system is as young as 125~Myr, i.e. zero-age main sequence, then obviously the isochrone-based ages reported in Table \ref{tab:params} are incorrect and some other parameters are systematically in error, since they are determined self-consistently with those ages.  While \teff\ and $R_*$ may be robust, $M_*$ and thus $\log g$ will be underestimated, and due to covariance between [Fe/H] and $\log g$ \citep{Torres2006}, metallicity is overestimated.

\primary\ is significantly older than most other dippers, which are typically found in $\lesssim10$~Myr-old star-forming regions, as well as the 30-50~Myr-old UX Ori-type variable RZ Piscium \citep{Punzi2018}.  It joins a small number of known main sequence systems and evolved stars that exhibit transient dimming, e.g. R Corona Borealis stars \citep{Montiel2018}, a white dwarf with disintegrating planetesimals \citep{Vanderburg2015}, stars with ``exocomets"  \citep{Rappaport2018,Ansdell2019}, and the enigmatic F dwarf KIC\,8462852 \citep{Boyajian2016}.  \primary\ also falls in a region of the Rosenberg-Hertzsprung-Russell diagram between the higher-mass Herbig Ae/Be UX Orionis stars and the lower-mass M-type dippers where young stars with transient dimming are uncommon, or at least rarely found.

\begin{figure}
	\includegraphics[width=\columnwidth]{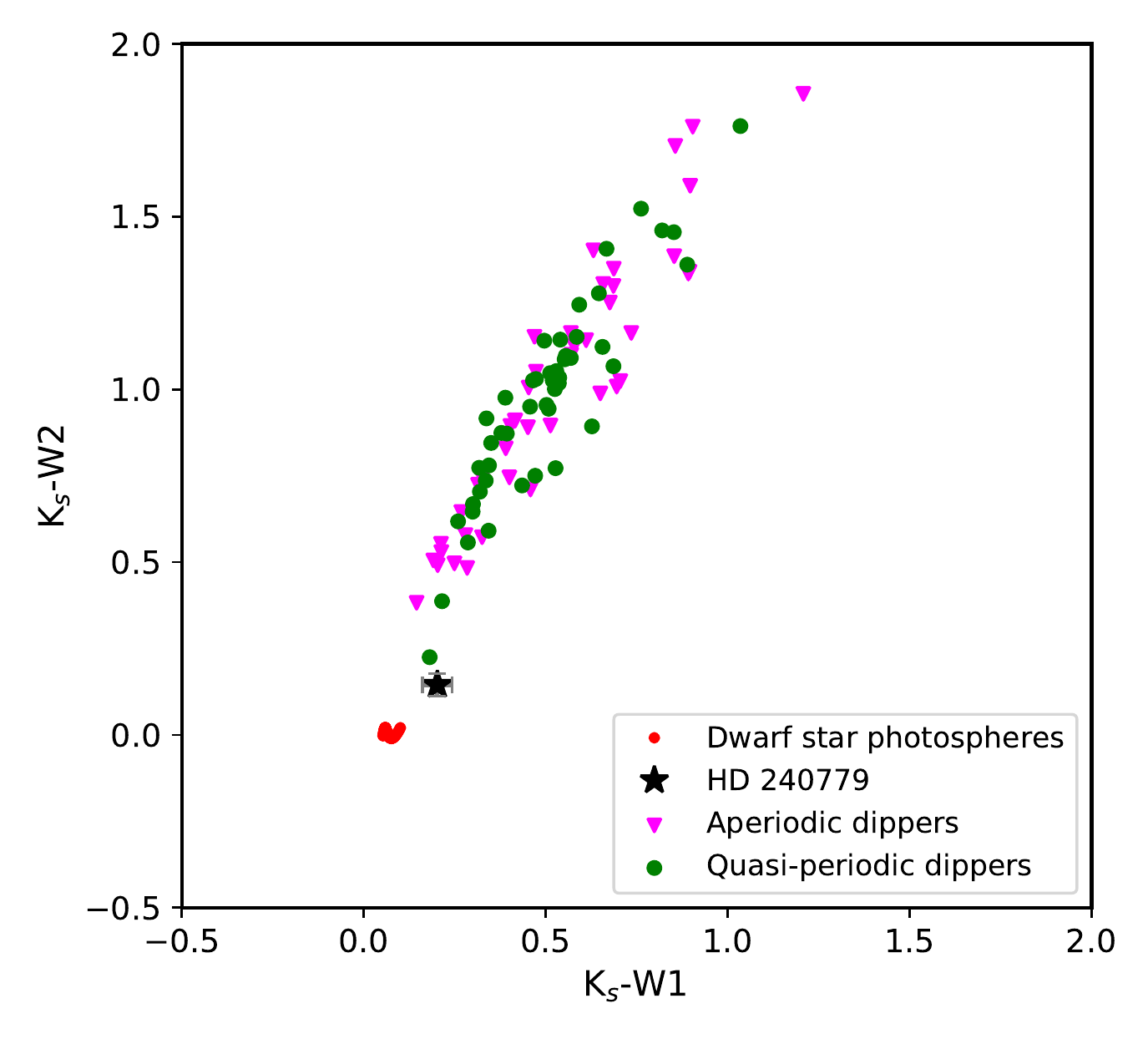}
    \caption{Excess infrared emission in the \wise\ W1 (3.4 $\mu$m) and W2 (4.6 $\mu$m) channels relative to 2MASS $K_s$ (2.2 $\mu$m), for \primary\ (shown as star symbol with error bars) relative to stars with aperiodic and quasi-periodic dipping identified by \citet{Cody2018} in Upper Scorpius and $\rho$ Ophiucus star-forming regions.  Also shown are the colors of photospheres of 3800-8300K dwarf stars from \citet{Jian2017}.}  
    \label{fig:excess}
\end{figure}

\primary, but not its K-type companion, has emission significantly in excess of the photosphere in the \emph{WISE} 12 and 22 $\mu$m channels indicative of circumstellar dust. In contrast to most young dipper stars, e.g., those observed by \ktwo\ in the 3-10~Myr-old $\rho$ Ophiuchus and Upper Scorpius star-forming regions \citep{Ansdell2016,Cody2018}, there is little or no excess emission at 3.4 and 4.6$\mu$m (Fig. \ref{fig:excess}), nor is there any emission in H$\alpha$ or the forbidden lines of oxygen indicative of accretion.  This indicates the absence of an inner disk, although this deficiency does not conflict with the amount of dust that is required to produce the dips alone, even if that dust is hot.  Assuming emission at the maximum (dust sublimation) temperature of 1500K, the radiation of the dust that produces a 10\% dip would be 0.7\% and 1\% of the photosphere at 3.4 and 4.6$\mu$m, respectively, and would be obscured by measurement error (Fig. \ref{fig:sed}).  

The excess emission at longer wavelengths resembles that of a blackbody at 470K and the inferred fractional infrared luminosity is $L_{\rm dust}/L_* \approx 2\times 10^{-3}$, within the range of ``debris disks" \citep{Hughes2018}.  The only dipper star in the \citet{Cody2018} sample that lacks significant excess emission at 3.4 and 4.6 $\mu$m is EPIC~205238942, an M0-type member of Upper Sco (confirmed by \gaia\ DR2 astrometry) with a strong rotational signal (period = 9.3 days) and stochastic dips of up to $\sim$15\%.  In these respects the stars are more ``evolved" relatives of RZ Piscium, a UX Ori-like star with much deeper dips (up to 2.5 magnitudes) and a much larger infrared excess ($L_{\rm dust} / L_* \approx 0.08$) that is adequately described by dust at a single (500K) temperature.

In this case, the lack of a substantial inner disk eliminates some of the proposed mechanisms involving accretion \citep{Bouvier2013} or vertical structures in a disk \citep{Ansdell2016}.  Impact disruption of asteroidal bodies or giant impacts during the final phase of rocky planet formation \citep{Morbidelli2016} have been invoked to explain warm dust around 30-50 Myr-old RZ Piscium, the 25 Myr-old $\beta$ Pictoris Moving Group member HD 172555 \citep{Johnson2012}, and ID8, a star with a rapidly time-varying excess in the $\sim$35~Myr cluster NGC\,2547 \citep{Su2019}.  Something analogous to these impact-driven disks might explain the excess infrared emission of 125 Myr-old \primary.  The dust required to produce the dips of \primary\ is equivalent to a 100-km size body that has been completely disrupted into 10 $\mu$m grains, and 470K corresponds to the temperature expected of dark grains on the orbit of Mercury.  The lifetime of dust against Poynting Robertson drag at this distance is $\sim10^3$yr, and thus the dust is ephemeral.  Warm, 12 \micron\ excess-producing dust is rare ($\sim$1\%) around young ($<$120~Myr) , non-T Tauri-like stars \citep{Kennedy2013}. which might explain why \primary\ has detectable dust but \secondary\ does not. 

\begin{figure}
	\includegraphics[width=\columnwidth]{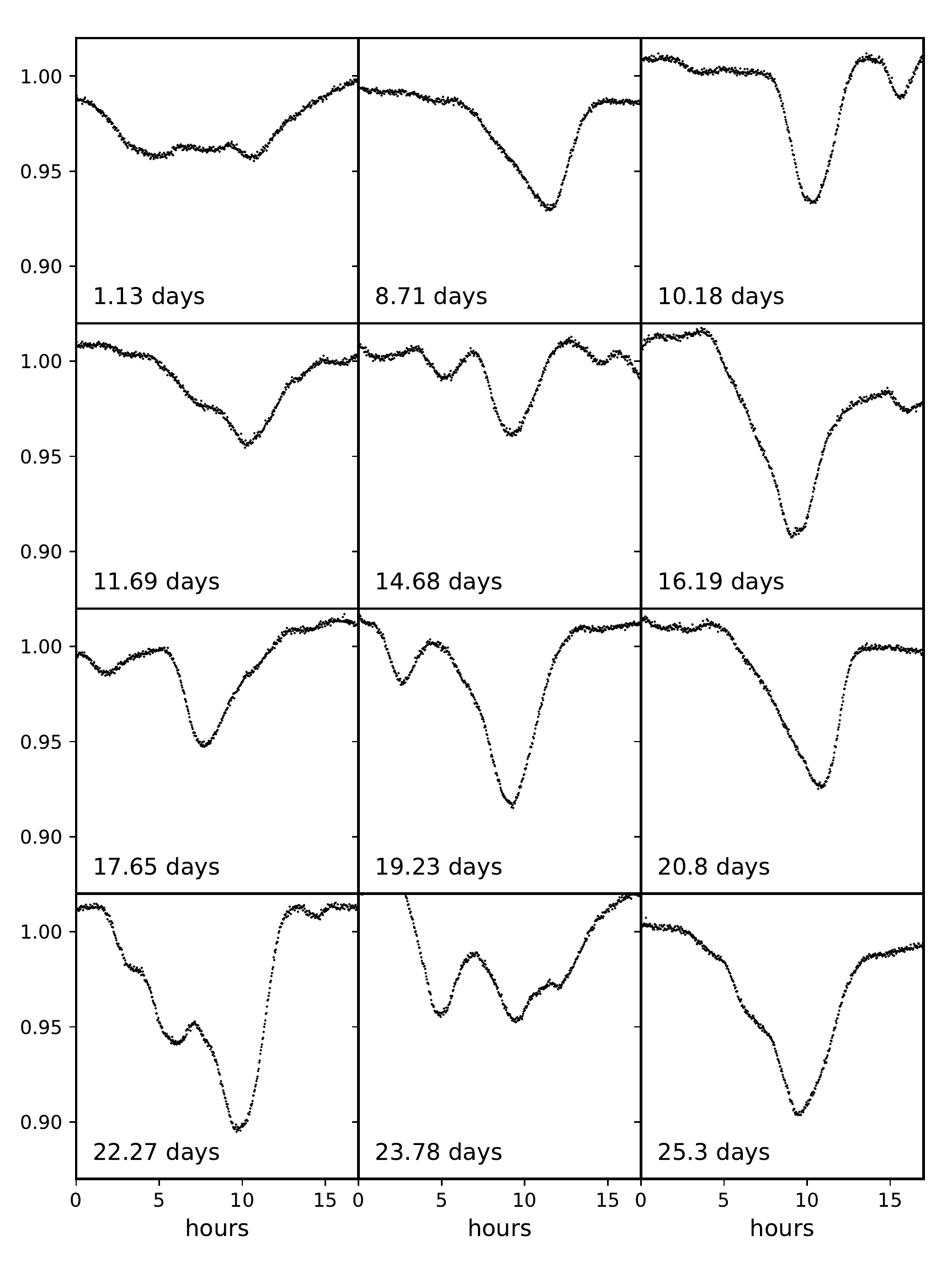}
    \caption{Gallery of the 12 deepest dips, in order of occurrence.}
    \label{fig:gallery}
\end{figure}

\begin{figure}
	\includegraphics[width=\columnwidth]{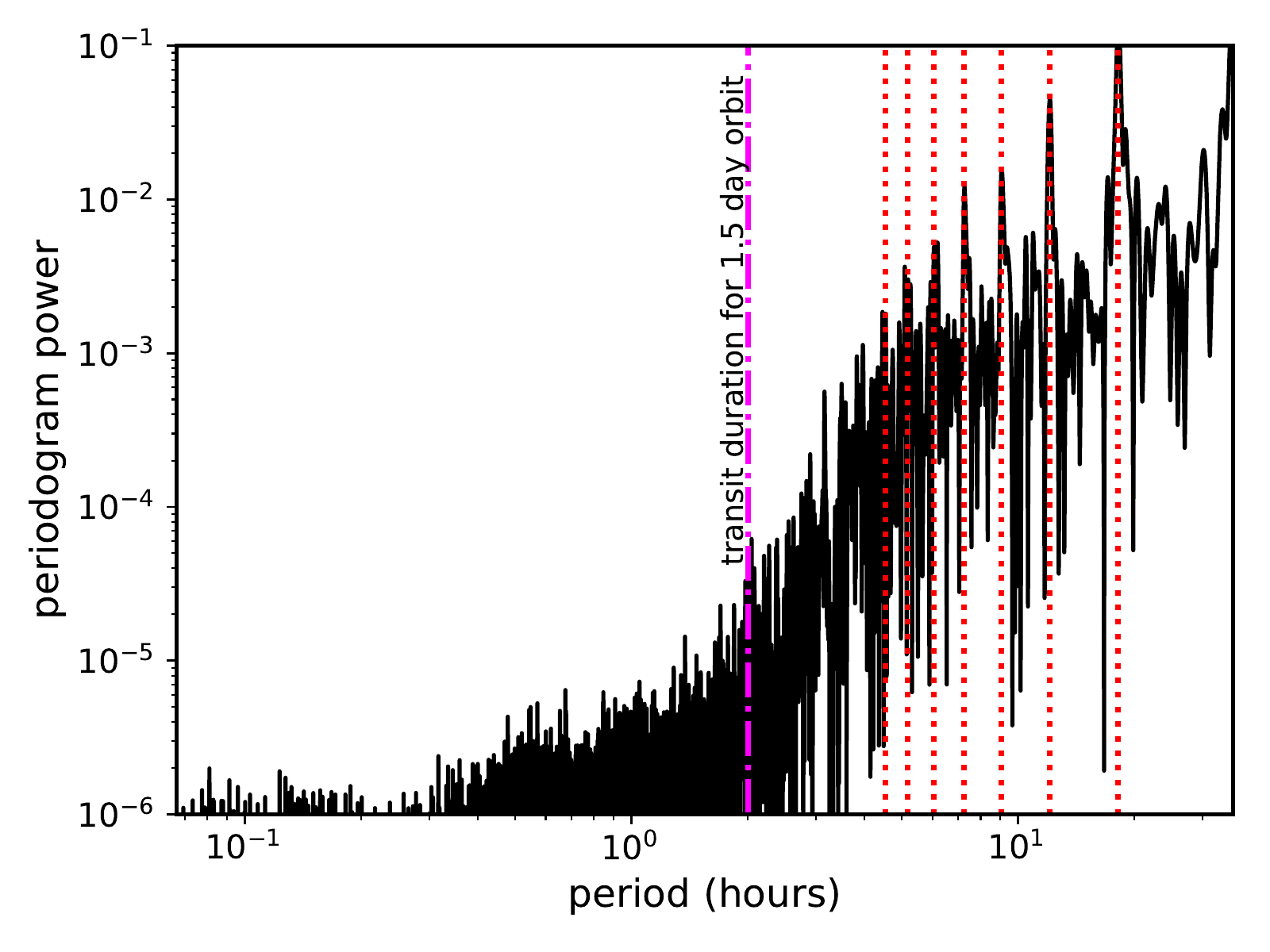}
    \caption{Periodogram of the \primary\ lightcurve between the Nyquist limit (4 min) and the 1.51-day period of the dip signal.  The expected 2-hour maximum transit duration of a small body on a circular 1.51 day around \primary\ is marked as the vertical magenta dash-dot line.  Vertical red dotted lines mark harmonics of 1.51 days.}
    \label{fig:power}
\end{figure}

This leaves the question of the nature and origin of the material responsible for the quasi-periodic signal at 1.51 days.  The phased signal (Fig. \ref{fig:phase} more closely resembles the periodic dippers found in Upper Sco \citep{Ansdell2016} and is more ephemeral and irregular than that of the ``evaporating" planets Kepler-1520b and K2-22b \citep{Rappaport2012,Sanchis-Ojeda2015}, suggesting a different mechanism.  If the period is that of a Keplerian orbit, the semi-major axis is about 6 stellar radii,  If we have correctly identified the stellar rotation period (4.2 days) then this material is orbiting within the stellar co-rotation radius.  The maximum transit duration of obscuring objects much smaller than the star and on a circular 1.51-day orbit is 2 hrs, thus a light curve should \emph{not} contain structure on much shorter timescales.  Fig. \ref{fig:gallery} illustrates the smoothness of the light curves of the largest dips at intervals less than a few hours, and Fig. \ref{fig:power} shows that the roll-off in periodogram power begins at around 5 hours and is complete at 2 hours.  The lack of structure in the light curve at intervals between 2 and 5 hours could simply mean that occulting dust is distributed or clustered on scales comparable to the stellar disk as well as eccentric orbits.  This is certainly suggested by the extended and composite dips in Figs. \ref{fig:tess_lc} and \ref{fig:gallery}.

The obscuring dust could arise from planetesimals that are disrupting due to tides or stellar irradiation.  For the Roche limit to extend to 6 stellar radii, the density of a strengthless (fluid) body would have to be $\le$0.1 g~cm$^{-1}$, well below that of rubble-pile asteroids or comets \citep{Carry2012}.  The equilibrium temperature of dark surfaces at this separation is $\approx$1700K, near or beyond the vaporization temperature of most solids. Thus the dust could arise from the irradiation-driven evaporation of rocky planetesimals, analogous to the ``evaporating planets" detected by \kepler\ and \ktwo\ \citep{Rappaport2012,Sanchis-Ojeda2015}.  The mass in the evaporative wind has to be at least comparable to that in the dust.  The mass in micron-size grains required to occult 10\% of the disk of the star is about $4 \times 10^{14}$ kg, equivalent to a body several km across.  Given the rapid changes in dip shape, the dispersal time must be of order a few orbits;  replenishment by evaporation would require $\sim 10^{15}-10^{16}$~W, far more than available by irradiation of such a body.  However, energy-limited evaporation of a larger ($\gtrsim$100~km) body could supply the material, albeit only for a few decades to centuries before its complete destruction.

Although the dust producing the dips is not responsible for the \emph{WISE}-detected infrared excess, the fact that these two short-lived phenomena occur simultaneously around \primary\ -- but not \secondary\ -- suggests that they are causally related.  One scenario is that a perturber has excited planetesimals from a disk onto eccentric, crossing orbits where they collide and produce the dust at $\sim$0.5~AU and bring them close to \primary, where their orbits can be circularized by non-gravitational forces.  The stellar companion \secondary\ could be this perturber, especially if the periastron is as small as 100~AU (Fig. \ref{fig:periastra}).  If a primordial disk is inclined by more than 40 deg. to the plane of the binary orbit, the Kozai-Lidov mechanism can act to send planetesimals on highly eccentric, highly inclined orbits \citep{Naoz2016}.  For a stellar semi-major axis of 1000~AU, planetesimals on initially circular orbits as close as a few AU and as far as a few tens of AU will experience Kozai-Lidov oscillations and could be placed on ``star-grazing" orbits at the present epoch; for objects on smaller orbits, post-Newtonian precession dampens the oscillation; the shorter oscillation times for planetesimals on wider orbits mean that they would have already been destroyed \citep{Naoz2016}.  

Future work could include sub-mm observations of \primary\ to constrain the flux at longer wavelengths and possibly to resolve any outer disk and determine its inclination, precise RV measurements to obtain the acceleration of the stars and better constrain the orbit and hence periapsis and the dynamical impact on the disk, and observations of dips at multiple wavelength to detect the expected wavelength-dependent scattering indicative of dust.

\section*{Acknowledgements}

The authors thank Jonathan Gagn\'{e} for discussion of young moving group identifications.  TJ and DL gratefully acknowledge Allan R. Schmitt for making his light curve-examining software LcTools freely available.  This paper includes data collected by the \tess\ mission and archived by the Mikulski Archive for Space Telescopes at the Space Telescope Science Institute. Funding for \tess\ is provided by the NASA Explorer Program. We acknowledge support by the NASA High-End Computing (HEC) Program through the NASA Advanced Supercomputing Division at Ames Research Center for the production of the SPOC data products. Some of the data presented herein were obtained at the W. M. Keck Observatory, which is operated as a scientific partnership among the California Institute of Technology, the University of California and the National Aeronautics and Space Administration.   The Observatory was made possible by the generous financial support of the W. M. Keck Foundation.  This research has made use of the NASA/ IPAC Infrared Science Archive, which is operated by the Jet Propulsion Laboratory, California Institute of Technology, under contract with the National Aeronautics and Space Administration. 





\bsp	
\label{lastpage}
\end{document}